\documentclass[3p,times]{elsarticle}

\usepackage{ecrc}

\volume{00}

\firstpage{1}

\journalname{Computer Networks}

\runauth{Avrachenkov et al.}

\jid{COMNNET}

\jnltitlelogo{Computer Networks}

\usepackage{amssymb}
\usepackage[figuresright]{rotating}

\usepackage{amsmath,amsfonts}
\usepackage{amsthm}
\usepackage{color}

\usepackage{caption}
\usepackage{subcaption}
\usepackage{graphicx}

\usepackage{pjnotation}

\usepackage{pjthm}

\usepackage{pjannotate}

\usepackage[all]{xy}

\usepackage{booktabs}

\input pjlatex.def

\newtheorem{proposition}{Proposition}

\newtheorem{definition}{Definition}

\begin{document}

\begin{frontmatter}



\dochead{}

\title{Congestion Control of TCP Flows in Internet Routers \\ by Means of Index Policy\tnoteref{ack}}
\tnotetext[ack]{The authors are grateful to
the two anonymous referees for valuable suggestions that helped to improve the presentation
of the paper. This research was initiated during the stay of J. Doncel at BCAM during May-September 2011 funded by the BCAM
Internship program. This research was partially supported by grant MTM2010-17405 (Ministerio de Ciencia e Innovaci\'on, Spain).}


\author{K. Avrachenkov$^{a}$, U. Ayesta$^{b,d,e}$,
J. Doncel$^{c,e,g}$, P. Jacko$^{f,g}$  \\ \ \\
$^{a}$INRIA Sophia Antipolis, France \\
$^b$IKERBASQUE, Basque Foundation for Science, Bilbao, Spain\\
$^c$Universite de Toulouse, INSA, LAAS, Toulouse, France\\
$^d$Universite de Toulouse, LAAS, Toulouse, France\\
$^e$LAAS-CNRS, Toulouse, France\\
$^f$Department of Management Science \& LANCS Initiative, Lancaster University, Lancaster, UK\\
$^g$BCAM-Basque Center for Applied Mathematics, Bilbao, Spain
}

\address{}

\begin{abstract}
\noindent In this paper we address the problem of fast and fair transmission
of flows in a router, which is a fundamental issue in networks like
the Internet. We model the interaction between a source using the
Transmission Control Protocol (TCP) and a
bottleneck router with the objective of designing optimal packet
admission controls in the router queue. We focus on the relaxed
version of the problem obtained by relaxing the fixed buffer
capacity constraint that must be satisfied at all time epoch. The
relaxation allows us to reduce the multi-flow problem into a family
of single-flow problems, for which we can analyze both theoretically
and numerically the existence of optimal control policies of special
structure. In particular, we show that for a variety of parameters,
TCP flows can be optimally controlled in routers by so-called index
policies, but not always by threshold policies. We have also
implemented the index policy in Network Simulator-3 and tested in a
simple topology their applicability in real networks. The simulation
results show that the index policy achieves a wide range of
desirable properties with respect to fairness between
different TCP versions, across users with different
round-trip-time and minimum buffer required to achieve full
utility of the queue.
\end{abstract}


\end{frontmatter}

% \linenumbers

% The body of the paper starts here:
\section{Introduction}
This paper deals with congestion control and buffer management, two
of the most classical research problems in networking. The objective
of congestion control is to control the traffic injected in the
network in order to avoid congestive collapse. Most traffic in the
Internet is governed by TCP/IP (Transmission Control Protocol and
Internet Protocol) (\cite{rfc2581,Jac88}). TCP protocol tries to
adjust the sending rate of a source to match the available bandwidth
along the path. In the absence of congestion signals from the
network TCP increases congestion window gradually in time, and upon
the reception of a congestion signal TCP reduces the congestion
window, typically by a multiplicative factor. Buffer management
determines how congestion signals are generated. Congestion signals
can be either packet losses or ECN (Explicit Congestion
Notifications) (\cite{rfc3168}). At the present state of the Internet,
nearly all congestion signals are generated by packet losses.
Packets can be dropped either when the router buffer is full or when
an AQM (Active Queue Management) scheme is employed (\cite{FJ93}).

In this paper we develop a rigorous mathematical framework to model
the interaction between a TCP source and a bottleneck queue with the
objective of designing optimal packet admission controls in the
bottleneck queue. The TCP sources follow the general family of
\emph{Additive Increase Multiplicative Decrease} pattern that TCP
versions like New Reno or SACK follow (\cite{FF96}). A TCP
source is thus characterized by the decrease factor, which
determines the multiplicative decrease of the congestion window in the event
of a congestion notification (in TCP New Reno the decrease factor takes the value 1/2).

The objective is to design a packet admission control strategy to use the resources efficiently and provide satisfactory user
experience. Mathematically we formulate the problem using the Markov Decision Process (MDP) (\cite{Puterman2005}) as a dynamic
resource allocation problem, which extends the restless bandit model introduced in \cite{Whittle1988}. The router aims at
maximizing the total aggregated utility. As utility function we adopt the parameterized family of generalized $\alpha$-fair
utility functions, which depending on the value of $\alpha$ permit to recover a wide variety of utilities such as max-min,
maximum throughput and proportional fair (\cite{MW00,AltmanEtal2008}). The fixed bandwidth capacity constraint that must be
satisfied at all time epochs makes that the problem can be solved analytically only in simplistic scenarios. However, the
problem becomes tractable if the fixed capacity constraint is relaxed so that the bandwidth allocation must be satisfied only
on average, known as the Whittle relaxation. This relaxation allows to see congestion control at the router as a family of
per-flow admission control problems, thus reducing the complexity of a multi-flow problem into a family of single-flow
problems. In our main contribution, for the single-flow problem we analyze both theoretically and numerically the existence of
optimal control policies of special structure. In particular, we show that for a variety of parameters, TCP flows can be
optimally controlled in routers by so-called index policy, but not always by
threshold policy. %We also develop an algorithm that computes the
%value of the indices.

This solution approach based on the Whittle relaxation has gained notorious
success in recent years and has been for instance used in scheduling in
wireless networks, dynamic/stochastic knapsack problems, online
marketing, or military applications. We refer to \cite{GGW11} for a recent account
on the methodology and applications. The interest in the approach
is justified by mathematical results that show that, under some
additional assumptions, the heuristic is asymptotically optimal
(\cite{WW90}). In practice it has been reported on various occasions
that the heuristic provides a close-to-optimal solution to the
problem studied \cite[Chapter 6]{GGW11}.

The index policy requires that the TCP sources are assigned a dynamic index that depends on their current congestion window and
the TCP variant implemented at the source. Such a \emph{transmission index} measures the efficiency of
transmission, and therefore can be interpreted as a priority for transmission. When a packet arrives and the router wants to
send a congestion notification (i.e., the buffer is full or the implemented AQM requires to drop or mark a packet), the router
will choose the packet with \emph{smallest index}. In the event that more than one packet share the same smallest index, the
packet that has been the longest in the queue is chosen.

Building on the cooperative nature of the TCP protocol, we assume that, apart from adjusting the window
according to the congestion signals received from the network, the source writes the current index into the packet header,
which are read by the routers. Alternatively, index policy can be implemented only in the routers, which however requires to
identify or infer the flows, their TCP variants, and the current congestion window of each flow.

Note that our scheme requires as much or less
information exchanges than the related congestion control protocols
XCP (\cite{KatabiEtal2002}), ACP (\cite{DukkipatiEtal2005}) and RCP
(\cite{LestasEtal2008}). Moreover, these congestion control protocols
aim at achieving max-min fairness, whereas our scheme allows to
achieve several notions of fairness by using the generalized
$\alpha$-fairness framework. Max-min is a particular
case of $\alpha$-fairness when the parameter $\alpha$ goes to
infinity. The most important TCP variants are FAST TCP (\cite{WJLH06}),
TCP CUBIC (\cite{HRX08}) and TCP Compound (\cite{TSZS06}), which are
implemented in Linux and Windows, respectively. The main originality
of our approach with respect to previously existing variants is that
our index protocol aims at achieving efficiency and fairness on
shorter-term time scales, by maximizing the mean fairness of the actual rate
instead of the fairness of the mean rates.
%and not only in the long-run.

We have implemented our solution in NS-3 (\cite{ns3}) and performed
extensive simulations in a benchmark topology to explore and
validate the properties of the algorithm, and to assess the
improvement with respect to \emph{DropTail router} (that drops
packets when the buffer is full) and \emph{RED router} (that has
implemented the Random Early Detection (RED) AQM mechanism). In the simulations we
focus on the case $ \alpha = 1 $, since it was shown that the
current Internet (with DropTail routers) maximizes the aggregate sum of
logarithmic utilities of the time-average transmission rates
(\cite{Kel97}).

The simulation results show that the algorithm has several desirable
properties with respect to fairness and efficiency:
\begin{itemize}
\item We achieve inter protocol fairness between different versions of AIMD TCP.
\item We achieve fairness with respect to the round-trip-time on the scale of congestion periods.
\item We improve fairness across users with different round-trip-time.
\item We improve fairness across users compared to DropTail and RED.
\item In a router implementing index policies, a smaller buffer size is needed to get the same throughput.
\end{itemize}

We believe that our approach opens a new avenue of research to
design in a combined manner congestion control and queue management
policies. This paper represents a first attempt, and briefly discuss in
\autoref{sec:discussion} how to assess several aspects that we did not
consider in this paper, as, for instance, the performance of the algorithm in the
presence of short-lived TCP connections, or when implemented in many routers in the network. We also
believe that we provide fundamental framework and ideas useful or
directly applicable also in next-generation Internet architectures
such as the Information-centric networking (ICN) and wireless networks.

The rest of the paper is organized as follows. In \autoref{sec:relatedwork} we put
our work in the context of the existing literature. In \autoref{sec:problemsection}
we describe the model and in \autoref{sec:MDP} we state the problem. Section V shows the
benefits of the relaxation of the described problem. In \autoref{sec:singleflow} we analyze
the index policies for single-flow problems. We also establish several
properties of index policies to be taken into account. With this results in the hand,
we present in \autoref{sec:simulation} the conclusion of the simulation done with NS-3.
In \autoref{sec:discussion} we discuss possible modeling extensions and practical implementation
aspects. Finally, some conclusions are drawn and possible extensions are discussed in
\autoref{sec:conclusion}.

A three-pages extended abstract of this paper appears in \cite{AADJ12}.

\section{Related Work on TCP and buffer management}
\label{sec:relatedwork}

The seminal work by \cite{CK74} established the basis
of TCP, and \cite{Jac88} added several key features and
brought the TCP protocol very close to its current form. TCP is a
completely distributed algorithm run by the end hosts, and it aims
to share the resources of the network among all flows in an
efficient and fair way. In other words, assuming that the network is
responsible for the end-to-end transmission of packets, TCP tries to
determine the fair share of the connections. The latter is
characterized by the \emph{congestion window}, denoted by $cwnd$,
which captures the number of packets a connection can have
simultaneously in the network at any given time. The basic
principles according to which TCP works are extremely simple. The
key idea is the dynamic window sizing scheme proposed by
\cite{Jac88}. Consider that any packet has an identifier that allows
the receiver to identify it uniquely. The sender sends packets to
the destination, and the destination sends back to the sender small
packets (known as acknowledgements, or simply ACKs) acknowledging
the correct reception of each packet. This simple mechanism allows
the sender to infer whether its packets are reaching the
destination, as well as implicitly to infer the congestion
level of the network. While packets reach the destination, the
sender is allowed to increase the amount of packets per unit of time
it injects into the network. More precisely, TCP increases by one
the value of $cwnd$ in the absence of network congestion notification.

On the other
hand, when a packet of a sequence of packets is lost or dropped, the
receiver will notice it by receiving an out-of-order packet, and it
will send a special acknowledgment (known
as duplicate ACK) back to the source. Upon the reception of three
duplicate ACKs or an ECN acknowledgement (in case of marked packet),
the source will infer that the network is congested and it will take
two actions: first, the lost packet (and the subsequent ones already sent) will be
retransmitted and second, the packet sending rate will be reduced
by a multiplicative factor. In the last two decades there has been a
huge effort in order to improve and develop new congestion control
algorithms, and we briefly mentioned some of them in the
introduction. An extensive overview and comparison of different TCP
version is given in \cite{LLS05}.

Buffer management algorithms determine how packets are dropped/marked, or
more generally how congestion signals are generated. Packets can be
dropped either when the router buffer is full, which is referred to
as DropTail, or when AQM (Active Queue Management) scheme, e.g., RED, is
employed (\cite{FJ93}). At the present state of the Internet, nearly
all congestion signals are generated by packet drops. An
alternative to packet drops is ECN (Explicit Congestion
Notifications) (\cite{rfc3168}), which marks packets but does not drop
them. Despite the tremendous research effort it seems that given the
ambiguity in the choice of parameters, in reality AQM schemes are
rarely used in practice. On the other hand, in the basic Drop Tail
routers, the buffer size is the only one parameter to tune apart
from the router capacity. We refer the interested reader to \cite{Vu07} and
references therein for more information on the problem of optimal
choice of buffer size. The main drawbacks of the basic DropTail
mechanism is the synchronization of TCP flows and RTT unfairness.
TCP connections with shorter RTT achieve faster high transmission
rates than TCP connections with larger RTT.

Our approach to AQM, based on sound theoretical background of Markov Decision Processes, does not require tuning of any
additional parameters and allows to realize the following interesting features: higher network utilization and more fair data
transmission of users with different TCP variants, different RTTs .

%This makes the choice of the router
%buffer size a very important consideration in the TCP/IP network
%design, and therefore the study of the optimal buffer size has
%attracted lot of attention. There are some empirical rules for the
%choice of the router buffer size. The first proposed rule of thumb
%for the choice of the router buffer size was to choose the buffer
%size equal to the BDP (Bandwidth-Delay Product) of the outgoing link
%\cite{VS94}. This recommendation is based on very approximative
%considerations and it can be justified only when a router is
%saturated with a single long-lived TCP connection. In a relevant
%paper the authors of \cite{AKM04} suggest that the minimal buffer
%size for the full system utilization should be chosen inversely
%proportional to the quare root of the number of competing
%connections. We refer the interested reader to \cite{Vu07} and
%references therein for more information on the problem of optimal
%choice of buffer size.

%\newpage
\section{Problem description}
\label{sec:problemsection}
We describe in this section the congestion control problem of
multiple flows at a bottleneck router. Suppose there are $ K $ flows
trying to deliver packets to their destinations via a bottleneck
router with the following parameters:
\begin{itemize}
\item $ C $ the bandwidth, i.e., the deterministic \emph{link capacity} (in packets per second);

\item $ B $ the buffer size (in packets);
\end{itemize}

Suppose further that flow $ k \in \setK := \{ 1, 2, \dots, K \} $
has implemented an additive-increase/multiplicative-decrease (AIMD)
mechanism as in the Transmission Control Protocol (TCP). The
congestion window $cwnd$ is adapted according to received
acknowledgements: for each received non-duplicate acknowledgment
(\emph{positive acknowledgement}), $cwnd$ is increased by the
reciprocal of the current value of $cwnd$ (which approximately
corresponds to an increase by one packet during a round-trip-time
RTT without lost packets) unless it has reached the maximum
advertised congestion window allowed $ N_{ k } $
\begin{align}
cwnd := \min \left\{ \left\lfloor cwnd + \frac{ 1 }{ cwnd } \right\rfloor, N_{ k } \right\};
\end{align}
for each triple duplicate acknowledgment (\emph{negative
acknowledgment}), $cwnd$ is decreased multiplicatively using the
formula
\begin{align}
cwnd := \max \left\{ \left\lfloor \gamma_{ k } \cdot \text{congestion window} \right\rfloor, 1 \right\},
\end{align}
where $ 0 \le \gamma_{ k } < 1 $ is the multiplicative decrease
factor. The function $\lfloor \cdot \rfloor$ denotes the floor
function. We consider that independently of the number of dropped/marked packets,
the congestion window is only decreased once per RTT. The
flow starts in the (deterministic) initial congestion window value $
n_{ k } $ and it always has packets willing to deliver.
For model transparency we ignore timeouts (and the corresponding slow-start
phase of TCP), since they are known to be significantly less frequent than
congestion events and we expect they would not affect our solution
significantly. Nevertheless, it would be possible to deal with it
with our approach.
%http://userver.ftw.at/~ricciato/darwin/itc05.pdf

The objective for the router is to use the resources efficiently and
provide satisfactory user experience so that
\begin{itemize}

\item the overall number of delivered packets per long time intervals (time-average throughput) is as large as possible and

\item the flows are treated fairly by having their congestion windows (number of packets in the network) as equal as
    possible and

\item the utilization of the bottleneck queue is as high as possible

\end{itemize}

\section{Markov Decision Processes Model}
\label{sec:MDP}
In this section we present a general mathematical formulation of
the congestion control problem using a Markov Decision Process (MDP)
framework of resource allocation (\citep{Jacko2009met}), which extends
the restless bandit model (\citep{Whittle1988}). To the best of our
knowledge restless bandit model formulation has not been used
previously in congestion control, an area in which an important body
of literature is devoted to deterministic fluid models (\cite{SS07}).

In order to make the model tractable and to be able to design in
 a simple, implementable solution, we will consider several
simplifications of the features of the problem. Nevertheless, the
performance of the solution is then evaluated in
Section~\ref{sec:simulation} in the original setting of the problem
without simplifications described above.

Let us consider the time slotted into discrete time epochs $ t \in
\setT := \{ 0, 1, 2, \dots \} $, which correspond to time periods of
one round-trip time (RTT), which is assumed equal for all the flows.
However, in Section \ref{sec:simulation} this constrain disappears and
we obtain nice results too. So, we consider users with equal RTT in the
theoretical part of this paper, although we can conclude from the simulations
this constrain can be omitted.

We assume that all packets are of the same size, which we further
define to be one bandwidth capacity unit. The router takes decisions
about admitting or rejecting the flows at every time epoch $ t $.

\begin{figure}[t]
\annotate{0.001\columnwidth}{0.08\columnwidth}{$X_{1} (t)$}
\annotate{0.001\columnwidth}{0.19\columnwidth}{$X_{2} (t)$}
\annotate{0.001\columnwidth}{0.31\columnwidth}{$X_{3} (t)$}
\annotate{0.001\columnwidth}{0.507\columnwidth}{$X_{K} (t)$}
\annotate{0.14\columnwidth}{0.102\columnwidth}{$W_{1, X_{1} (t)}^{\text{sent}}$}
\annotate{0.14\columnwidth}{0.219\columnwidth}{$W_{2, X_{2} (t)}^{\text{sent}}$}
\annotate{0.14\columnwidth}{0.334\columnwidth}{$W_{3, X_{3} (t)}^{\text{sent}}$}
\annotate{0.14\columnwidth}{0.537\columnwidth}{$W_{K, X_{K} (t)}^{\text{sent}}$}
\annotate{0.32\columnwidth}{0.07\columnwidth}{$a_{1} (t)$}
\annotate{0.32\columnwidth}{0.19\columnwidth}{$a_{2} (t)$}
\annotate{0.32\columnwidth}{0.3\columnwidth}{$a_{3} (t)$}
\annotate{0.31\columnwidth}{0.51\columnwidth}{$a_{K} (t)$}
\annotate{0.43\columnwidth}{0.19\columnwidth}{$W_{1, X_{1} (t)}^{a_{1} (t)}$}
\annotate{0.43\columnwidth}{0.457\columnwidth}{$W_{K, X_{K} (t)}^{a_{K} (t)}$}
\annotate{0.61\columnwidth}{0.285\columnwidth}{$B (t)$}
\annotate{0.77\columnwidth}{0.09\columnwidth}{$R_{1, X_{1} (t)}^{a_{1} (t)}$}
\annotate{0.77\columnwidth}{0.2\columnwidth}{$R_{2, X_{2} (t)}^{a_{2} (t)}$}
\annotate{0.77\columnwidth}{0.32\columnwidth}{$R_{3, X_{3} (t)}^{a_{3} (t)}$}
\annotate{0.77\columnwidth}{0.517\columnwidth}{$R_{K, X_{K} (t)}^{a_{K} (t)}$}
\includegraphics[width=\columnwidth, clip=true, trim=0pt 70pt 0pt 0pt]{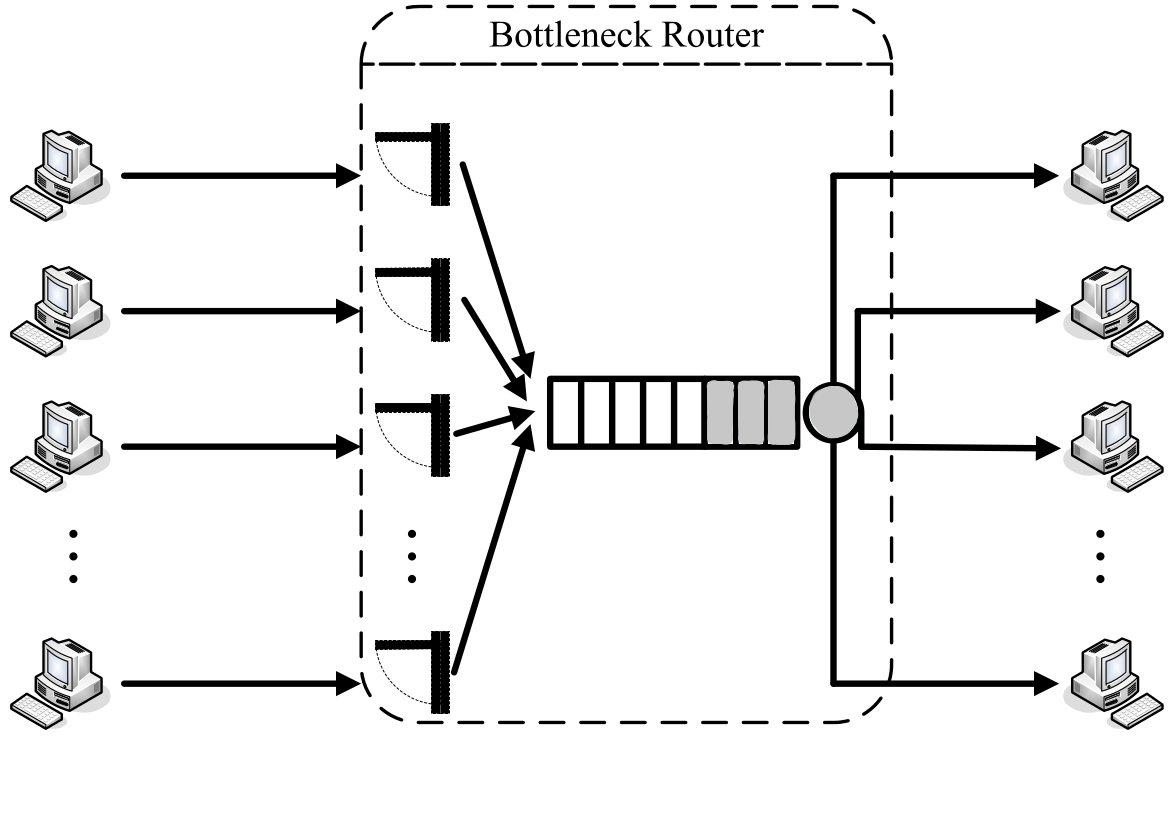}%
\caption{A scheme of $K$ flows sharing a bottleneck router.}\label{fig:network2}
\end{figure}

\subsection{AIMD Flows}
Every flow can be allocated either the capacity required by its
current congestion window (being \emph{admitted}) and transmitted,
or zero capacity (being \emph{rejected}). We denote by $ \setA := \{
0, 1 \} $ the \emph{action space}, where $ 0 $ corresponds to
blocking and $ 1 $ corresponds to admitting. This action space is
the same for every flow $ k $.

Each flow $ k $ is defined independently of other flows as the tuple
\begin{align*}
\left( \setN_{ k }, \left( \vecimmedW_{ k }^{ a } \right)_{ a \in \setA }, \left( \vecimmedR_{ k }^{ a } \right)_{ a \in \setA }, \left( \matP_{ k }^{ a } \right)_{ a \in \setA } \right),
\end{align*}
where
\begin{itemize}
\item $ \setN_{ k } := \{ 1, 2, \dots, N_{ k } \} $ is the \emph{state space}, i.e., a set of possible congestion windows
    flow $ k $ can set;

\item $ \vecimmedW_{ k }^{ a } := \left( \immedW_{ k, n }^{ a } \right)_{ n \in \setN_{ k } } $, where $ \immedW_{ k, n }^{
    a } $ is the expected one-period capacity consumption (in
    number of packets), or \emph{work} required by flow $ k $ at
    state $ n $ if action $ a $ is decided at the beginning of a
    period; in particular, $ \immedW_{ k, n }^{ 0 } := 0 $ and $
    \immedW_{ k, n }^{ 1 } := n $;

\item $ \vecimmedR_{ k }^{ a } := \left( \immedR_{ k, n }^{ a } \right)_{ n \in \setN_{ k } } $, where $ \immedR_{ k, n }^{
    a } $ is the expected one-period generalized $ \alpha $-fairness or \emph{reward} earned by flow $ k $ at state $ n $
    if action $ a $ is decided at the beginning of a period; in particular $ \immedR_{ k, n }^{ 0 } := 0 $ and
\begin{align*}
     \immedR_{ k, n }^{ 1 } :=
     \begin{cases}
       \displaystyle \frac{ ( 1 + n )^{ 1 - \alpha } - 1 }{ 1 - \alpha }, & \text{ if } \alpha \neq 1, \\
       \log ( 1 + n ), & \text{ if } \alpha = 1;
     \end{cases}
\end{align*}

\item $ \matP_{ k }^{ a } := \left( p_{ k, n, m }^{ a } \right)_{ n, m \in \setN_{ k } } $ is the flow-$ k $ stationary
    one-period \emph{state-transition probability matrix} if action $ a $ is decided at the beginning of a period, i.e., $
    p_{ k, n, m }^{ a } $ is the probability of moving to state $ m $ from state $ n $ under action $ a $; in particular, $
    p_{ k, n, m }^{ 1 } = 1 $ iff $ m = \min \left\{ n + 1 , N_{ k } \right\} $ (representing additive increase) and $ p_{
    k, n, m }^{ 0 } = 1 $ iff $ m = \max \left\{ \left\lfloor \gamma_{ k } \cdot n \right\rfloor, 1 \right\} $
    (representing multiplicative decrease); the remaining probabilities are zero.

\end{itemize}

The dynamics of flow $ k $ is thus captured by the \emph{state
process} $ X_{ k } ( \cdot ) $ and the \emph{action process} $ a_{ k
} ( \cdot ) $, which correspond to state $ X_{ k } ( t ) \in \setN_{
k } $ and action $ a_{ k } ( t ) \in \setA $, respectively, at all
time epochs $ t \in \setT $.  The states $n \in \setN_{ k } $ denote
possible levels of the sending rate. In particular
$\immedW_{n}^{\text{sent}} := n $  can therefore be interpreted as
the bandwidth capacity the flow requires for complete transmission
at the current period.

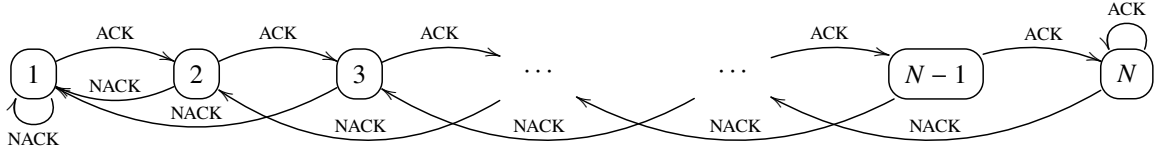
\begin{figure*}[t]
\begin{displaymath}
\xymatrixcolsep+{20pt}
\entrymodifiers={++[F-:<7pt>]}
\xymatrix{
  1 \ar@/^10pt/[r]^-{\text{ACK}} \ar@(dr,dl)^-{\text{NACK}}
& 2 \ar@/^10pt/[r]^-{\text{ACK}} \ar@/^10pt/[l]_-{\text{NACK}}
& 3 \ar@/^10pt/[r]^-{\text{ACK}} \ar@/^20pt/[ll]_-{\text{NACK}}
& *+++{\cdots} \ar@/^25pt/[ll]_-{\text{NACK}}
& *+++{\cdots} \ar@/^10pt/[r]^-{\text{ACK}} \ar@/^25pt/[ll]_-{\text{NACK}}
& N - 1 \ar@/^10pt/[r]^-{\text{ACK}} \ar@/^25pt/[ll]_-{\text{NACK}}
& N \ar@(ur,ul)_-{\text{ACK}} \ar@/^25pt/[ll]_-{\text{NACK}}
}
\end{displaymath}
\caption{A model of an AIMD flow as a Markov chain. The arrows
represent one-period transitions among the states $1, 2, \dots, N$
after a congestion-free (ACK) and a congestion-experienced (NACK)
transmission.}\label{fig:MC}
\end{figure*}

The schematic behavior of the AIMD flow as a Markov chain is shown
in Figure \ref{fig:MC}, where ``ACK'' represents a congestion-free
delivery of the flow packets to the receiver (positive
acknowledgments) and ``NACK'' represents a congestion-experienced
transmission (negative acknowledgments). Notice that the evolution
is completely deterministic; this can be extended to stochastic evolution
like for instance in \cite{JackoSanso2012restarting}.

\subsection{Multi-flow Optimization Problem}
The flows dynamics is as follows (see \autoref{fig:network2}). At
epoch $ t $, the sender of each flow $ k \in \setK $ sets its state
$ X_{ k } ( t ) $ (that depends on whether the previous-epoch
workload was transmitted, given by the positive or negative
acknowledgement of the receiver of flow $ k $ sent back to the
sender in the previous period) and sends the workload of $ \immedW_{
k, X_{ k } ( t ) }^{\text{sent}} $ packets to the bottleneck router.
However, only $ 0 \le \immedW_{ k, X_{ k } ( t ) }^{ a_{ k } ( t ) }
\le \immedW_{ k, X_{ k } ( t ) }^{\text{sent}} $ packets are allowed
to queue in the buffer (admitted) for being transmitted. The
transmitted packets arrive to the receiver of flow $ k $, who
obtains the fairness (reward) $ \immedR_{ k, X_{ k } ( t ) }^{ a_{ k
} ( t ) } $. If the router admitted the flow, then the receiver
sends positive acknowledgement back to the sender; otherwise
negative acknowledgment is sent. According to this, the sender sets
its next-epoch state $ X_{ k } ( t + 1 ) $ and repeats the process.

The congestion avoidance decision at the router is taken in the
following way. At epoch $ t $ the router observes the  states
(congestion windows) $ X_{ k } ( t ) $ of all flows $ k \in \setK $.
Based on that it decides the flow actions $ a_{ k } ( t ) $ (which
may be viewed to be taken in virtual gates, as illustrated in
\autoref{fig:network2}), instantaneously appends (in FIFO order) $
\immedW_{ k, X_{ k } ( t ) }^{ a_{ k } ( t ) } $ packets of each
flow $ k $ to the buffer, and transmits (in FIFO order) $ W $
packets (or all the packets if there are less than $ W $ packets in
the buffer) during the period.

To summarize, the senders make no decisions and therefore the flows
dynamics can be modeled as a Markov chain.

At this moment we present a generic formulation of the congestion
control optimization problem. Let $ \Pi $ be the set  of all
history-dependent randomized policies. Denote by the symbol $
\Expectation_{\vecn}^{\pi} $ the conditional expectation given that
the initial conditions are $ \vecn := \left( n_{ k } \right)_{ k \in
\setK } $, and the policy applied is $ \pi \in \Pi $. The router
controller's problem to solve is under the discounted criterion (if
$ \beta < 1 $)
\begin{align}
\max_{\pi \in \Pi} \Expectation_{\vecn}^{\pi} \left[ \sum_{t = 0}^{\infty} \sum_{ k \in \setK } \beta^{t} \immedR_{ k, X_{ k } ( t ) }^{a_{ k } ( t ) } \right] & \label{obj} \\
\llap{\text{subject to \hskip1em}} \Expectation_{\vecn}^{\pi} \left[ \sum_{t = 0}^{\infty} \sum_{ k \in \setK } \beta^{t} \immedW_{ k, X_{ k } ( t ) }^{ a_{ k } ( t ) } \right] &  \leq \frac{\overline{W}}{1 - \beta} \label{soft} \\
 \sum_{ k \in \setK } \immedW_{ k, X_{ k } ( t ) }^{ a_{ k } ( t ) } & \le B, \text{ for all } t \in \setT. \label{hard}
\end{align}
The problem can also be formulated under the time-average criterion.
It is known by \cite[Lemma 7.1.8]{Puterman2005} that the optimal policy
for the time-average criterion can be obtained by the limit $ \beta
\to 1 $ of the optimal policy for the discounted problem.

The virtual capacity seen as the target time-average router
throughput $ \overline{W} $ is equal to  the bandwidth-delay-product
$C \times RTT$, that is, the maximum number of packets that can be
served in one slot. We note that an analogous constraint
\eqref{soft} formulation was used in \cite{MaEtal2008}. To avoid
the trivial problem of underloaded router, we assume that for
policy $ \pi $ that always admits we have
\begin{align*}
\Expectation_{\vecn}^{\pi} \left[ \sum_{t = 0}^{\infty} \sum_{ k \in \setK } \beta^{t} \immedW_{ k, X_{ k } ( t ) }^{ \text{sent}} \right] &  > \frac{\overline{W}}{1 - \beta} \\
\end{align*}

\section{Decomposition of the Multiple-Flows Problem}
\label{sec:decomp} The problem \eqref{obj}--\eqref{hard} is
difficult to solve due to the sample path constraint \eqref{hard}.
One possibility for relaxing the problem is to assume that the
buffer space $ B $ is infinite, so that the constraint \eqref{hard}
is trivially fulfilled. Another possibility is to relax that
constraint as did \cite{Whittle1988}, by requiring it only in
discounted expectation, i.e.,
\begin{align*}
\Expectation_{\vecn}^{\pi} \left[ \sum_{t = 0}^{\infty} \sum_{ k \in
\setK } \beta^{t} \immedW_{ k, X_{ k } ( t ) }^{ a_{ k } ( t ) }
\right] &  \leq \frac{B}{1 - \beta}.
\end{align*}
However, such a constraint is weaker than \eqref{soft}, because $ B
\geq \overline{W} $, since in our model $B$ is the total number of
packets that the buffer can handle in one RTT, whereas
$\overline{W}$ corresponds to the number of packets that can be
served in one RTT.

Either of these two relaxation possibilities results in omitting the
constraint \eqref{hard}.  Thus, we end up with a problem formulation
\eqref{obj}--\eqref{soft}, which is analogous to the Whittle
relaxation of the multi-armed restless bandit problem
(\citep{Whittle1988}).

The standard solution of such a formulation is by solving for each $ \nu $ the Lagrangian relaxation of
\eqref{obj}--\eqref{soft}, which is
\begin{align}
\max_{\pi \in \Pi} \Expectation_{\vecn}^{\pi} \left[ \sum_{t = 0}^{\infty} \sum_{ k \in \setK } \beta^{t} \left( \immedR_{ k, X_{ k } ( t ) }^{a_{ k } ( t ) } - \nu  \immedW_{ k, X_{ k } ( t ) }^{a_{ k } ( t ) } \right)    \right] + \nu \frac{\overline{W}}{1-\beta} \label{lagr}
\end{align}
where $ \nu $ is the Lagrangian parameter that can be interpreted as a per-packet
\emph{transmission cost}. The Lagrangian theory assures that there exists $ \nu^{*} $,
for which the Lagrangian relaxation \eqref{lagr} achieves optimum of
\eqref{obj}--\eqref{soft}. Since for any fixed $ \nu $ the flows are independent and the second
term of \eqref{lagr} is constant, we can decompose \eqref{lagr} into $ K $ individual-flow
problems.

\begin{proposition}
\label{prop:decomposition} Let $ \Pi_{k} $ be the set of all
history-dependent randomized policies for flow $ k $, and
individual-flow policies $ \pi_{k}^{*} \in \Pi_{k} $ such that they
form the joint policy $ \pi^{*} \in \Pi $. If for a given parameter
$ \nu $, each policy $ \pi_{k}^{*} $ for $ k \in \setK $ optimizes
the individual-flow problem
\begin{align}
\max_{\pi_{k} \in \Pi_{k}} \Expectation_{n_{k}}^{\pi_{k}} \left[ \sum_{t = 0}^{\infty} \beta^{t} \left( \immedR_{k, X_{k} (t)}^{a_{k} (t)} - \nu \immedW_{k, X_{k} (t)}^{a_{k} (t)} \right) \right], \label{individual}
\end{align}
then $ \pi^{*} $ optimizes the multi-flow problem \eqref{lagr}.
\end{proposition}

In \autoref{sec:singleflow} we will find  an optimal solution to
such a $ \nu $-parameter problem  in terms of flow- and
state-dependent \emph{Whittle indices} $ \nu_{k,n} $, which in our
setting can be interpreted as \emph{transmission indices}. If the
optimal transmission cost $ \nu^{*} $ is known, then these indices
define the following optimal policy for problem
\eqref{obj}--\eqref{soft}: ``At each time slot admit all the flows
of actual-state transmission index greater than the transmission
cost $ \nu^{*} $ and reject the remaining flows''.

Since in practice $ \nu^{*} $ is typically unknown, the buffer space
is finite, and it is desirable to have work-conserving transmission
in order to increase bandwidth utilization, we use the transmission
indices to define practically feasible policy for
problem~(\ref{obj})-(\ref{hard}):

\emph{Heuristic Policy:} In every slot order the flows in decreasing
order with respect to their current indices, and admit the flows
until reaching the constraint~(\ref{hard}).

Following the results in the literature, we expect that our
heuristic policy will be asymptotically optimal as the number of
flows and buffer capacity grow to infinity (\cite{GGW11}).

In Section~\ref{sec:simulation} we will develop a heuristic
admission control policy for a TCP/IP network at packet level and
study its performance with NS-3 simulations. In this case the index
of each packet represents the admission priority into the buffer. If
the buffer is full, the packet in the queue with smallest
transmission index will be dropped.

\section{Index Policies for Single-Flow Subproblems}
\label{sec:singleflow}

The single flow admission control subproblem can be optimally solved
by means of index policies under some index-existence conditions
(\citep{Nino2007top}). In a particular case, the optimal policy is of
threshold type (if it is optimal to reject the flow under certain
congestion window, then it is optimal to do the same also under any
higher congestion window). However, depending on the parameters,
threshold policies may not be optimal, as will be illustrated in
~\ref{subsec:flows}. Therefore, in the rest of the paper we will obtain index
policies only numerically, and the investigation whether index
policies can be characterized by index values in closed form is left
out of this paper. (This is indeed possible in some cases, see, e.g.
\cite{JackoSanso2012restarting,AvrachenkovJacko2010tcp}).

Since no decisions are taken by the sender, congestion control is
implemented in the router. The router decides whether the incoming
flow in state $ n_{ k } $ should be admitted (and transmitted)
(achieved by action $ a_{ k } ( t ) = 1 $ of transmitting $
\immedW_{n_{ k }}^{1} := \immedW_{n_{ k }}^{\text{sent}}$ packets),
or rejected (action $ a_{ k } ( t ) = 0 $ of transmitting $ 0 $
packets). The difference of receiver rewards and transmission costs
(i.e., $\immedR_{n_{ k }}^{0} - \nu \immedW_{n_{ k }}^{0}$ and
$\immedR_{n_{ k }}^{1} - \nu \immedW_{n_{ k }}^{1}$) will be
henceforth called the \emph{net reward} under transmission cost
$\nu$.

To summarize, the (unconstrained) MDP problem of the single AIMD
flow addressed in this section under both the discounted and
time-average criteria is defined as follows:
\begin{itemize}
\item \emph{State space} is $ \setN_{ k } $.

\item \emph{Actions}: \emph{admitting} \emph{rejecting} are available in each state.

\item \emph{Dynamics if admitting}: If the flow is in state $n_{ k }$ and the flow is admitted at a given period, then during
    that period it generates net reward $\immedR_{n_{ k }}^{1} - \nu \immedW_{n_{ k }}^{1}$ and the flow moves to state
    $n_{ k } + 1$ (or remains in $ N_{ k } $, if $ n_{ k } = N_{ k } $) for the next period.

\item \emph{Dynamics if rejecting}: If the flow is in state $n_{ k }$ and the flow is rejected at a given period, then during
    that period it generates net reward $\immedR_{n_{ k }}^{0} - \nu \immedW_{n_{ k }}^{0}$ and the flow moves to state
    $n_{ k } - 1$ (or remains in $ n_{ k } $ if $ n_{ k } = 1 $) for the next period.
\end{itemize}

To evaluate a policy $\pi$ under the $\beta$-discounted criterion, we consider the following two measures. Let $\displaystyle
\totalW_{i}^{\pi} := \Expectation_{i}^{\pi} \left[ \sum_{t = 0}^{\infty} \beta^{t} \immedW_{X(t)}^{a(t)} \right] $ be the
\emph{expected total $\beta$-discounted bandwidth utilization} if starting from state $i$ under policy $\pi$. For convenience,
we will also call $\totalW_{i}^{\pi}$ the expected total $\beta$-discounted work, since the bandwidth utilization can be seen
as the work performed by the router in order to transmit the flow. Analogously we denote by $\displaystyle \totalR_{i}^{\pi} :=
\Expectation_{i}^{\pi} \left[ \sum_{t = 0}^{\infty} \beta^{t} \immedR_{X(t)}^{a(t)} \right] $ the \emph{expected total
$\beta$-discounted reward} if starting from state $i$ under policy $\pi$.

The objective \eqref{individual} is for each transmission cost
$\nu$,
\begin{align}
\max_{\pi \in \Pi} \totalR_{i}^{\pi} - \nu \totalW_{i}^{\pi}. \label{4:objective}
\end{align}

We will address this problem in the following subsections, noting that an optimal solution for the time-average variant is
obtained in the limit $ \beta \to 1 $.

\subsection{Threshold Policies and Indexability}

Since for finite-state MDPs there exists an optimal stationary policy independent of the initial state (\citep{Puterman2005}), we
narrow our focus only to those policies and represent them via \emph{admission sets} $\setS \subseteq \setN$. In other words, a
policy $\setS$ prescribes to admit the flow in states in $\setS$ and to reject the flow in states in $\setS^{\complement} :=
\setN \setminus \setS$.

We will therefore write $\totalR_{i}^{\setS}$ and $\totalW_{i}^{\setS}$ for the expected total $\beta$-discounted reward and
work, respectively, under policy $\setS$ starting from initial state $i$. Then, formulated for initial state $i$, the
optimization problem is the following combinatorial problem
\begin{equation}
\max_{\setS \subseteq \setN} \totalR_{i}^{\setS} - \nu \totalW_{i}^{\setS}.\label{problem}
\end{equation}

In order to solve this problem, we will be interested in two related structural properties, which are \emph{optimality of
threshold policies} and \emph{indexability}, formally defined below.

\begin{definition}[Optimality of Threshold Policies]
We say that problem \eqref{problem} is \emph{optimally solvable by threshold policies}, if for every real-valued $ \nu $ there
exists threshold state $ n \in \setN_{ k } \cup \{ 0 \} $ such that threshold policy admitting the flow in states $ \setS_{ N_{
k } }^{ ( n ) } := \{ m \in \setN_{ k } : m \le n \} $ and rejecting otherwise is optimal for problem \eqref{problem}.
\end{definition}

Of our interest will be the index proposed in
\cite{Whittle1988}, which often furnishes a nearly-optimal solution,
and typically recovers the optimal index rule if such exists. We
adopt the definition of indexability from \cite{Jacko2010jobs}.

\begin{definition}[Indexability]
We say that $ \nu $-parameter problem \eqref{problem} is \emph{indexable}, if there exist unique values $ -\infty
\le \nu_{ k, n } \le \infty $  for all $ n \in \setN_{ k } $ such that the following holds for every state $ n \in \setN_{ k }
$:
\begin{enumerate}

\item if $ \nu_{ k, n } \ge \nu $, then it is optimal to admit flow $ k $ in state $ n $, and

\item if $ \nu_{ k, n } \le \nu $, then it is optimal to reject flow $ k $ in state $ n $.

\end{enumerate}
The function $ n \mapsto \nu_{ k, n } $ is called the \emph{(Whittle) index}, and $ \nu_{ k, n } $'s are called the
\emph{(Whittle) index values}.
\end{definition}

An immediate consequence of the two definitions is formulated in the following previously known result.

\begin{proposition}
If problem \eqref{problem} is indexable and the index is nonincreasing, i.e., $ \nu_{ k, 1 } \ge \nu_{ k, 2 }
\ge \dots \ge \nu_{ k, N_{ k } } $, then problem \eqref{problem} is optimally solvable by threshold policies.
Moreover, for a given $ \nu $ the optimal threshold policy is $ \setS_{ N_{ k } }^{ n^{ * } } $ with $ n^{ * } \in \setN_{ k }
\cup \{ 0 \} $ such that $ \nu_{ k, n^{ * } } \ge \nu \ge \nu_{ k, n^{ * } + 1 } $ (defining $ \nu_{ k, 0 } := -\infty,
\nu_{ k, N_{ k } + 1 } := \infty $).
\end{proposition}

For transparency, we limit ourselves in this paper to numerical
testing of indexability and computation of the index values, which
then allow to conclude about optimality of threshold policies. We
have employed an algorithm based on the restless bandit framework
(\citep{Nino2007top}), which both numerically checks the conditions of
existence and calculates the index values, if they exist. It is a
one-run algorithm (analogous to parametric simplex method) and in
each step it calculates one of the index values. Thus, it performs $
N_{ k } $ steps, and the overall computational complexity of the
algorithm is $ O ( N_{ k }^{ 4 } ) $.

We have performed testing of indexability over a large number of
flows with different parameters. The algorithm always confirmed that
the flow was indexable. These tests give us evidence to conjecture
that the flows as defined in this paper are always indexable.
However, the complexity of the problem impeded us to find a
structure that could be exploited for establishing indexability
analytically. We will illustrate the difficulty in the next
subsections.

\subsection{One-, Two-, and Three-State Flows}
\label{subsec:flows}

In this subsection we provide analytical results showing that
optimality of threshold policies depends on values of parameters $
\alpha $, $ \beta $, $ N_{ k } $, and $ \gamma_{ k } $.

It is obvious ( see, e.g., \cite{JackoSanso2012restarting}) that a one-state flow (i.e.,
$ N_{ k } = 1 $) is indexable and solvable by threshold policies.
The index value for the unique state is $ \nu_{ k, 1 } = \immedR_{
k, 1 } / \immedW_{ k, 1 } $. Similarly, \cite{JackoSanso2012restarting}
proved that a two-state flow (i.e., $ N_{ k } = 2 $) is indexable
and solvable by threshold policies. The index values for the two
states are
\begin{align*}
\nu_{ k, 1 } &= \frac{ \immedR_{ k, 1 } }{ \immedW_{ k, 1 } }, &
\nu_{ k, 2 } &= \frac{ \immedR_{ k, 2 } + \beta ( \immedR_{ k, 2 } - \immedR_{ k, 1 } ) }{ \immedW_{ k, 2 } + \beta ( \immedW_{ k, 2 } - \immedW_{ k, 1 } ) }.
\end{align*}

Finally, \cite{JackoSanso2012restarting} proved that a three-state flow
(i.e., $ N_{ k } = 3 $) is indexable and solvable by threshold
policies, if
\begin{align*}
\matP_{ k }^{ 0 } = \begin{pmatrix} 1 & 0 & 0 \\ 1 & 0 & 0 \\ 1 & 0 & 0 \end{pmatrix},
\end{align*}
which holds if and only if $ \gamma_{ k } < 2/3 $. Under such a condition, the index values for the three states are
\begin{align*}
\nu_{ k, 1 } &= \frac{ \immedR_{ k, 1 } }{ \immedW_{ k, 1 } }, \\
\nu_{ k, 2 } &= \frac{ \immedR_{ k, 2 } + \beta ( \immedR_{ k, 2 } - \immedR_{ k, 1 } ) }{ \immedW_{ k, 2 } + \beta ( \immedW_{ k, 2 } - \immedW_{ k, 1 } ) }, \\
\nu_{ k, 3 } &= \frac{ \immedR_{ k, 3 } + \beta ( \immedR_{ k, 3 } - \immedR_{ k, 1 } ) + \beta^{ 2 } ( \immedR_{ k, 3 } - \immedR_{ k, 2 } ) }{ \immedW_{ k, 3 } + \beta ( \immedW_{ k, 3 } - \immedW_{ k, 1 } ) + \beta^{ 2 } ( \immedW_{ k, 3 } - \immedW_{ k, 2 } ) }.
\end{align*}
In all the above cases, the threshold policies are optimal, because the index values are nonincreasing (due to concavity of the
generalized $ \alpha $-fairness function).

It is tedious but straightforward to show (by a detailed inspection
of the algorithm that computes the indices) that $ \gamma_{ k } \ge
2/3 $ if and only if the index values for the three states are as
follows. If $ \alpha < 1 $, then
\begin{align*}
\nu_{ k, 1 } &= \frac{ \immedR_{ k, 1 } }{ \immedW_{ k, 1 } }, \\
\nu_{ k, 2 } &= \frac{ \immedR_{ k, 2 } - \beta \immedR_{ k, 1 } }{ \immedW_{ k, 2 } - \beta \immedW_{ k, 1 } }, \\
\nu_{ k, 3 } &= \frac{ \immedR_{ k, 3 } + \beta ( \immedR_{ k, 3 } - \immedR_{ k, 2 } ) }{ \immedW_{ k, 3 } + \beta ( \immedW_{ k, 3 } - \immedW_{ k, 2 } ) }.
\end{align*}
In this case, the threshold policies are optimal, because the index values are nonincreasing. On the other hand, if $ \alpha
\ge 1 $, then for some values of $ \beta $ the index values are as above, but for other values of $ \beta $ the index values
are
\begin{align*}
\nu_{ k, 1 } &= \frac{ \immedR_{ k, 1 } }{ \immedW_{ k, 1 } }, \\
\nu_{ k, 2 } &= \frac{ \immedR_{ k, 2 } + \beta ( \immedR_{ k, 3 } - \immedR_{ k, 1 } ) + \beta^{ 2 } ( \immedR_{ k, 3 } - \immedR_{ k, 2 } ) }{ \immedW_{ k, 2 } + \beta ( \immedW_{ k, 3 } - \immedW_{ k, 1 } ) + \beta^{ 2 } ( \immedW_{ k, 3 } - \immedW_{ k, 2 } ) }, \\
\nu_{ k, 3 } &= \frac{ \immedR_{ k, 3 } - \beta^{ 2 } \immedR_{ k, 1 } }{ \immedW_{ k, 3 } - \beta^{ 2 } \immedW_{ k, 1 } }.
\end{align*}
In this case, we have $ \nu_{ k, 1 } > \nu_{ k, 3 } > \nu_{ k, 2 } $, therefore threshold policies are not optimal in general.
In particular, threshold policy $ \setS_{ N_{ k } }^{ ( 2 ) } $ is never optimal.

\subsection{General Flows}

%MOVE THIS BEFORE? If a flow starts from the initial state $ n_{ k }
%= 1 $, then employing the optimal index policy always results in a
%threshold policy. This is due to the fact that the flow increases
%its congestion window additively, and so once a state whose index is
%below $ \nu $ is reached, the flow is rejected and changes its state
%to one with a higher index value (otherwise it would have been
%already rejected when passing through that state).

\begin{figure*}[th!]
\label{fig:indices}
\includegraphics[scale=0.55,clip=true,trim=90 290 80 290]{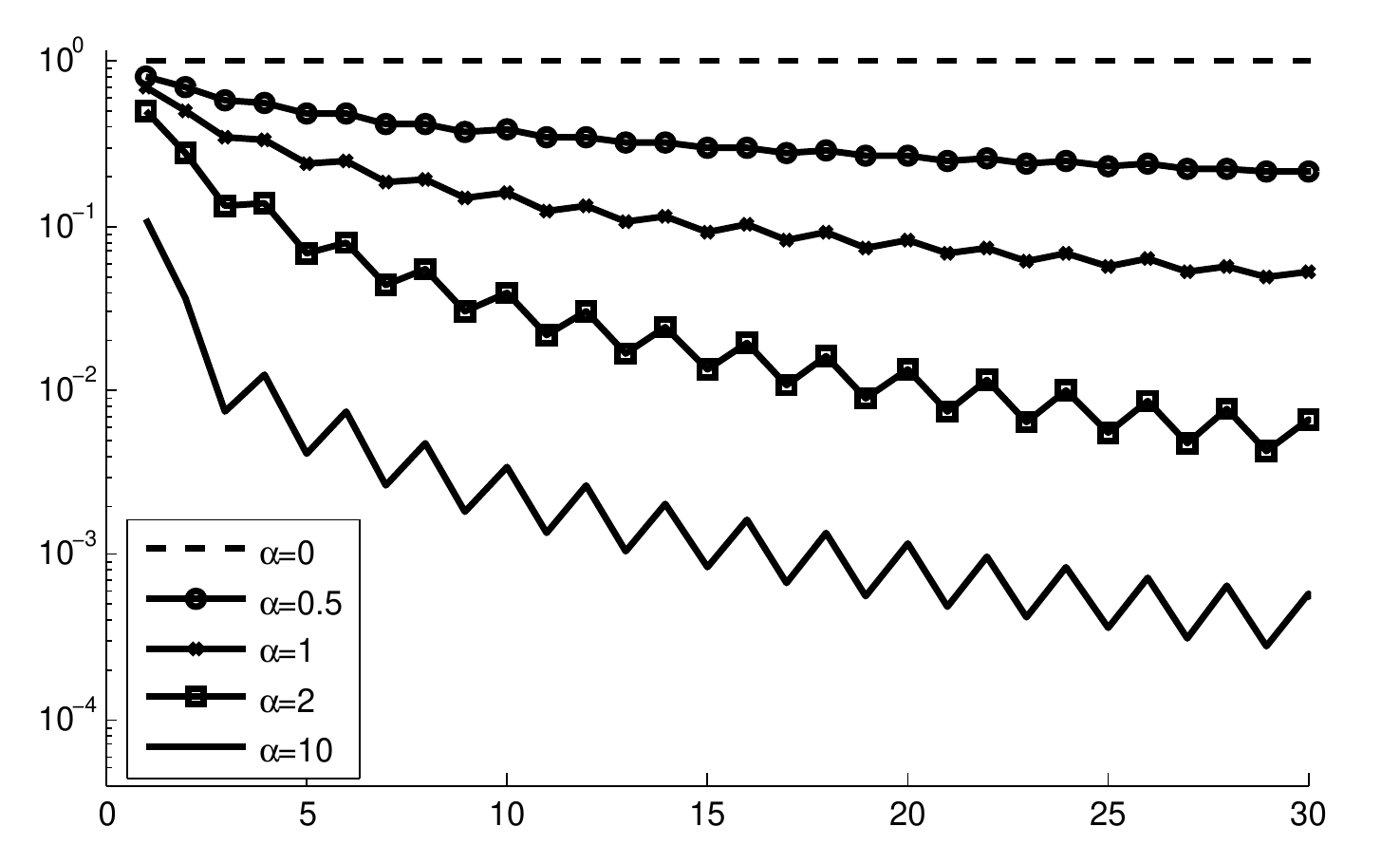}\hfill%   100 250 100 240        10 210 10 200
\includegraphics[scale=0.53,clip=true,trim=100 290 100 290]{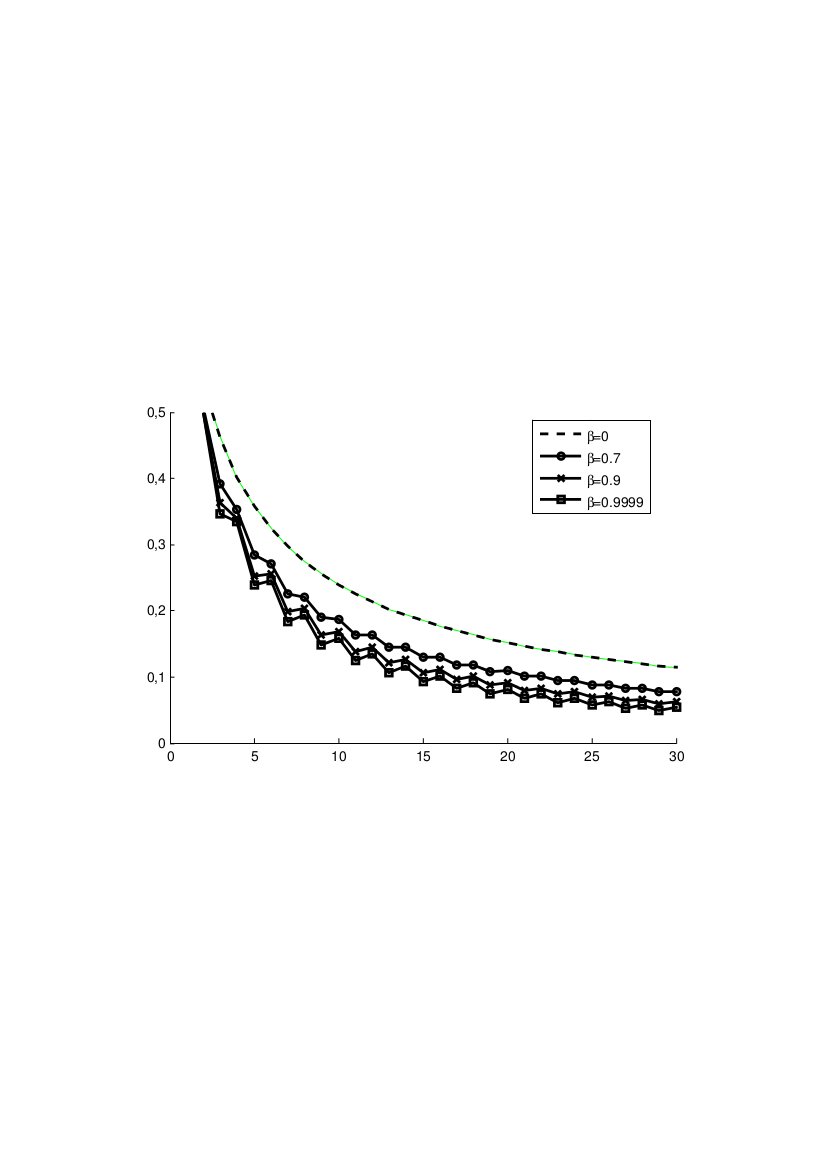}
\centerline{\footnotesize\hbox to 0.5\textwidth{\hfil (a) For different values of $ \alpha $ \hfil}\hbox to 0.5\textwidth{\hfil (b) For different values of $ \beta $ \hfil}}
%\includegraphics[width=0.5\textwidth,clip=true,trim=100 250 100 240]{figs/pic/varying_alpha_log.pdf}%
%\includegraphics[width=0.5\textwidth,clip=true,trim=100 250 100 240]{figs/pic/varying_beta_log.pdf}
%\centerline{\footnotesize\hbox to 0.5\textwidth{\hfil (c) Varying $ \beta $ (linear scale) \hfil}\hbox to 0.5\textwidth{\hfil (d) Varying $ \beta $ (log scale) \hfil}}
\includegraphics[scale=0.55,clip=true,trim=95 290 100 290]{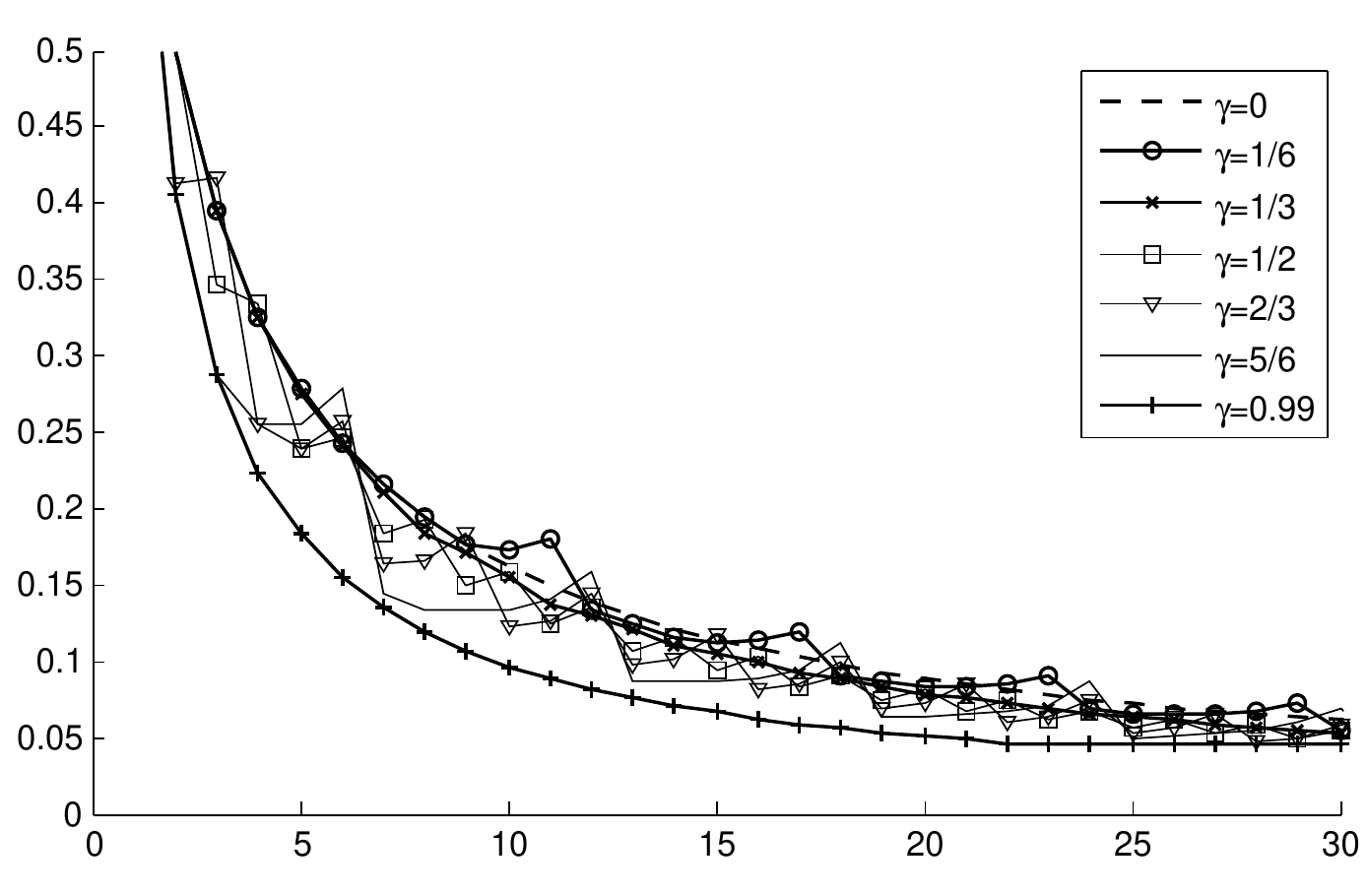}\hfill%
\includegraphics[scale=0.55,clip=true,trim=95 290 100 290]{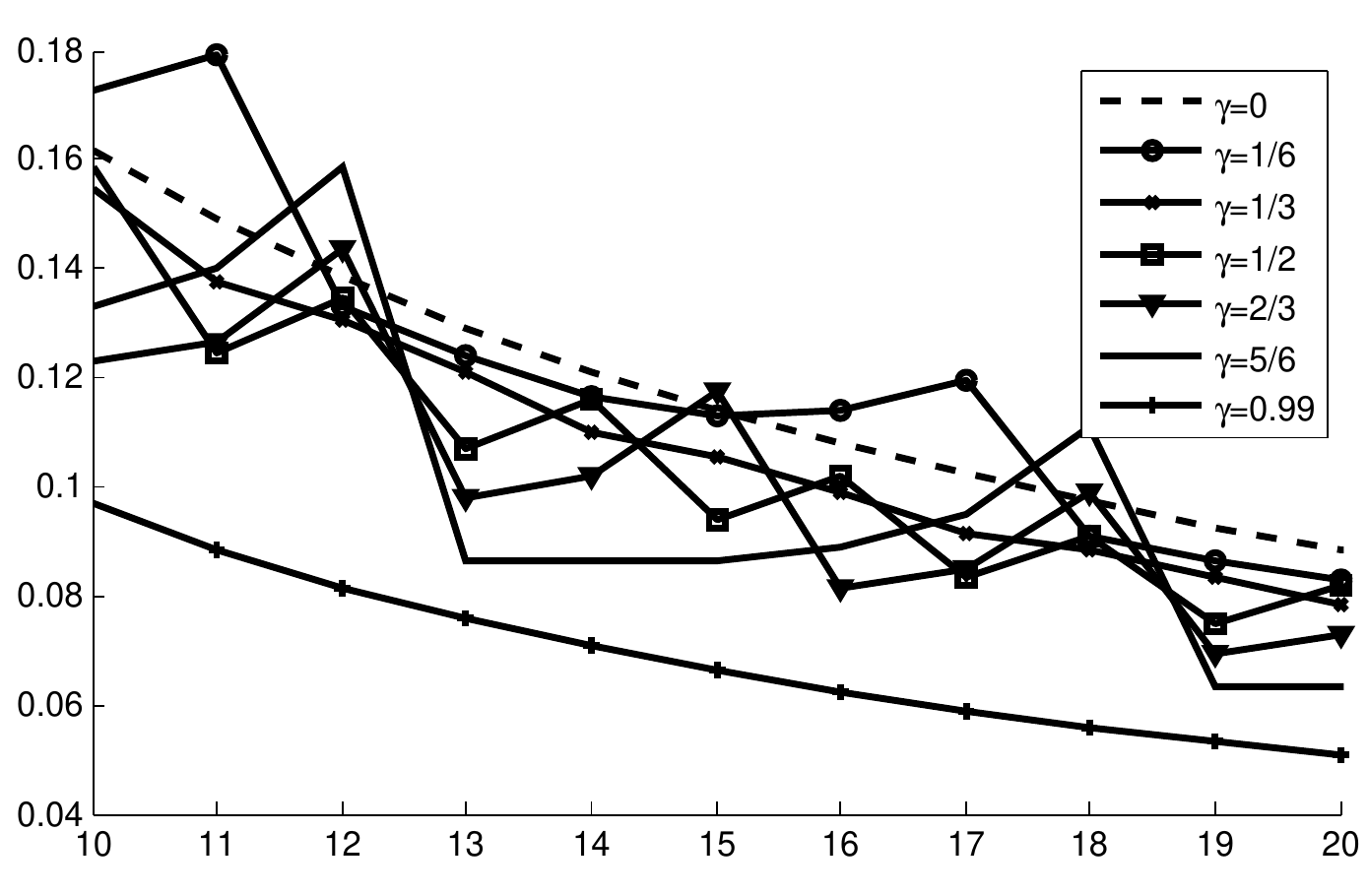}
\centerline{\footnotesize\hbox to 0.5\textwidth{\hfil (c) For different values of $ \gamma_{ k } $ \hfil}\hbox to 0.5\textwidth{\hfil (d) For different values of $ \gamma_{ k } $ (zoomed in) \hfil}}
\caption{Index values as a function of congestion window for different values of parameters.}\label{figures}
\end{figure*}

In this subsection we illustrate using numerical results that the
structure of optimal policies and the index values depend heavily on
values of parameters $ \alpha $, $ \beta $, $ N_{ k } $, and $
\gamma_{ k } $. This feature brings complexity to the mathematical
analysis of the problem, but on the other hand, it shows that
indices nicely capture the nature of different TCP variants under
different optimization criteria. Nevertheless, the figures provide
insights for better understanding of what an optimal admission
control policy is.

As for the basic instance, we set $ \alpha := 1 $, $ \beta := 0.9999
$, $ \gamma_{ k } := 1/2 $ and $ N_{ k } := 70 $. In
\autoref{figures} we present index values as a function of
congestion window for (a) different values of $ \alpha $, (b)
different values of $ \beta $, and (c)-(d) different values of $
\gamma_{ k } $. We note that we have observed that different values
of $ N_{ k } $ may also influence the index values, especially if
both $ \beta $ and $ \gamma_{ k } $ are large, but the differences
are often not noticeable in the figures, so they are omitted.

First of all, it can be seen in \autoref{figures}(a)-(b) that as $
\alpha $ or $ \beta $ diminish, threshold policies become optimal,
due to having the index values non-increasing, as opposed to the
non-monotone (zig-zag) index when both $ \alpha $ or $ \beta $ are
large. A similar behavior can be observed also for other values of $
\gamma_{ k } $, which are not reported in this paper.
\autoref{figures}(a) further indicates that index values are
decreasing in $ \alpha $ and their slopes for a given $ \alpha $ are
``more convex'', so higher $ \alpha $ ensures a stronger
discouragement of dropping packets belonging to flows with smaller
actual congestion window. Note that case $ \alpha = 0 $ gives all
the index values equal to $ 1 $, so no flow is prioritized over
other ones, which is analogous to the DropTail policy.
\autoref{figures}(b) indicates that index values are decreasing in $
\beta $, so longer flows get lower priority for transmission over
shorter flows with the same actual congestion window.

From \autoref{figures}(c)-(d) we can learn interesting insights as well. The index functions for $ \gamma_{ k } = 0, 0.99 $ are
smooth and non-increasing, as well as, rather surprisingly, $ \gamma_{ k } = 1/3 $. The remaining cases result in
``tooth-shaped'' index functions, with remarkably different tooth widths (akin to periods) of two ($ \gamma_{ k } = 1/2 $),
three ($ \gamma_{ k } = 2/3 $), and six ($ \gamma_{ k } = 1/6, 5/6 $ look similar, with their period point shifted), which may
happen because $ \gamma_{ k } $'s are multiples of six in this figure. In spite of that, index values of different $ \gamma_{ k
} $'s coincide or come very close to each other at congestion windows that are multiples of these periods. At multiples of six
(e.g., 6, 12, \dots), the index values are increasing in $ \gamma_{ k } $ (at least over the range $ 1/3, 1/2, 2/3, 5/6 $),
while at the subsequent points (e.g., 7, 13, \dots) the index values are decreasing in $ \gamma_{ k } $, so the priority
ordering of flows with such congestion windows is completely reversed.

We underline that \autoref{figures}(c)-(d) suggests that the index
function for $ \gamma_{ k } = 0 $ could be a reasonable smooth
approximation for the index functions for the remaining values,
especially for $ \gamma_{ k } = 1/3 $. This is of special interest
from practical point of view, because the index function for $
\gamma_{ k } = 0 $ is known in closed-form due to \cite{JackoSanso2012restarting}.
On the other hand, the same figure also suggests that the index
function for $ \gamma_{ k } = 0.99 $ (which is essentially a
birth-death dynamics) is a lower bound for the index functions for
the remaining values of $ \gamma_{ k } $, which is also known in
closed-form due to \cite{AvrachenkovJacko2010tcp}
. However, $
\gamma_{ k } = 0 $ does not necessarily provide an upper bound. Such
an upper bound could be obtained by using the myopic index by $
\beta = 0 $. Note also that as $ \beta $ gets closer to $ 0 $,
numerical differences between index values for different $ \gamma_{
k } $'s become smaller.

\begin{table*}[t]
\scriptsize
%\footnotesize
\begin{center}
\begin{tabular}{clllll}\\%ll}
\toprule
Scenario & Policy & Utilization & User fairness & Mean Queue Size & RTT \\
\midrule
0 & DropTail & 97.14\% & 0.999999 & 6.98 & 61.4\\
0 & RED &  97.63\% & 0.982572 & 6.11 & 58.77 \\
0 & Index & 98.22\% (+1.1\%, +0.6\%) & 0.999798 (-0.1\%, +1.8\%) & 8.07 & 64.7 (+5.2\%, +10.2\%) \\
\midrule
1 & DropTail & 81.16\% & 1.000000 & 2.41 & 47.4\\
1 & RED & 81.71\% & 0.998233 & 2.54  & 47.8\\
1 & Index & 88.45\% (+7.3\%, +6.7\%) & 0.999837 (-0.1\%, +0.1\%) & 2.89 & 48.8 (+2.3\%, +2\%) \\
\midrule
2 & DropTail & 90.84\% & 0.786535 & 7.1 & 61.8\\
2 & RED & 92.74\% & 0.911862 & 6.04 & 58.5\\
2 & Index & 95.37\% (+4.5\%, +2.6\%) & 0.962784 (+22.3\%, +5.5\%) & 7.34 & 62.5 (+1.1\%, 6.8\%)\\
\midrule
3 & DropTail & 86.36\% & 0.739875 & 6.63 & 60.3\\
3 & RED & 94.08\% & 0.821966 & 6.31 & 59.4\\
3 & Index & 96.86\% (+10.5\%, +2.8\%) & 0.917895 (+24.5\%, +11.6\%) & 6.84 & 62.5 (+3.6\%, +5.2\%)\\
\midrule
4 & DropTail & 93.84\% & 0.765318 & 6.11 & See \autoref{subsec:changing_prop_del}\\%138.8\\
4 & RED & 91.6\% & 0.89428 & 5.81 & \\%137.8\\
4 & Index & 94.97\% (+1.1\%, +3.3\%) & 0.929756 (+21.4\%, +3.9\%) & 6.37 & \\%141 (+1.5\%, +2.3\%)\\
\bottomrule
\end{tabular}
\end{center}
%\medskip
\caption{Utilization of the bottleneck queue, the Jain's fairness index between users
(over the $ 20 $ seconds interval), the Mean Queue Size (in number of packets) of the bottleneck buffer and the RTT of a connection of one user (in ms)
%and the Mean Waiting Time (in miliseconds) observed in the simulation in
% the Jain's fairness index between subintervals of $ 30 $ seconds, observed
in the simulations for
different scenarios. In parentheses, the improvement of index policy with respect to DropTail and RED.}\label{results}
\end{table*}

\section{Simulations results}
\label{sec:simulation}
In this section we present experimental results from implementing
index policy in simulations. We assume that TCP sources include in
the header of each packet the index value corresponding to the
actual $cwnd$. Based on the mathematical results we define the
following heuristic index policy to  be implemented in the Internet
routers:

\emph{Heuristic index policy at packet level:} Upon a packet
arrival, if the buffer is not full, then admit the packet.
Otherwise, drop the packet (either the new one or from the queue)
with smallest index value. In case of ties, drop the packet that has
been the longest in the queue.

We have employed and modified TCP New Reno in the NS-3 simulator (\cite{ns3}) to
obtain the results in several scenarios. The main objective of this
section is to show implementability of the proposed index policy, to
give fundamental insights about its effect, and to evaluate possible
gains of these results with the DropTail
and RED policies. As the measure of fairness we employ the Jain's fairness
index, whose value ranges from $ 1 $ (perfect fairness) to $ 1/K $
in a $ K $-user system (\cite{JCH94}).

We focus on case $ \alpha = 1 $, since it was shown that the loss
networks (like the current Internet with DropTail) maximizes the
aggregate sum of logarithmic utilities of the time-average
transmission rates (\cite{Kel97}). Note that our approach (with $
\alpha = 1 $) maximizes the aggregate sum of the time-average
logarithmic utilities of the immediate transmission rates.

The access links of each user is $ 5 $Mb/s and the delay is $10
$ms. The delay of the bottleneck link in this scenario is $ 10 $ms
and the bandwidth capacity of the bottleneck link is $ 1500 $kb/s.
The packet size is $ 576 $ Bytes. We set $\beta=0.9999$ which approximates the
time-average criterion. We set the maximum value of the congestion
window of each of the users to $ N_{ k } = 70 $, required to compute
the index values.

For each of the five scenarios below, we plot the time evolution
under the DropTail, RED and index policy of the following elements: the
size of the queue in the router buffer, the congestion window of
each of the users, and (only for index policy) the index value of
each of the users.

In all figures of the simulation section, user 1 is depicted with a
black and thin line, and user 2 with a black and thick solid line.

The results are summarized in \autoref{results}.

\subsection{Scenario 0 (Baseline): Two Symmetric Users}

\begin{figure}[t!]
\centering
        \begin{subfigure}[h!]{0.5\textwidth}
	  % \centering	
                \includegraphics[scale=0.55,clip=true, trim=105 275 105 270]{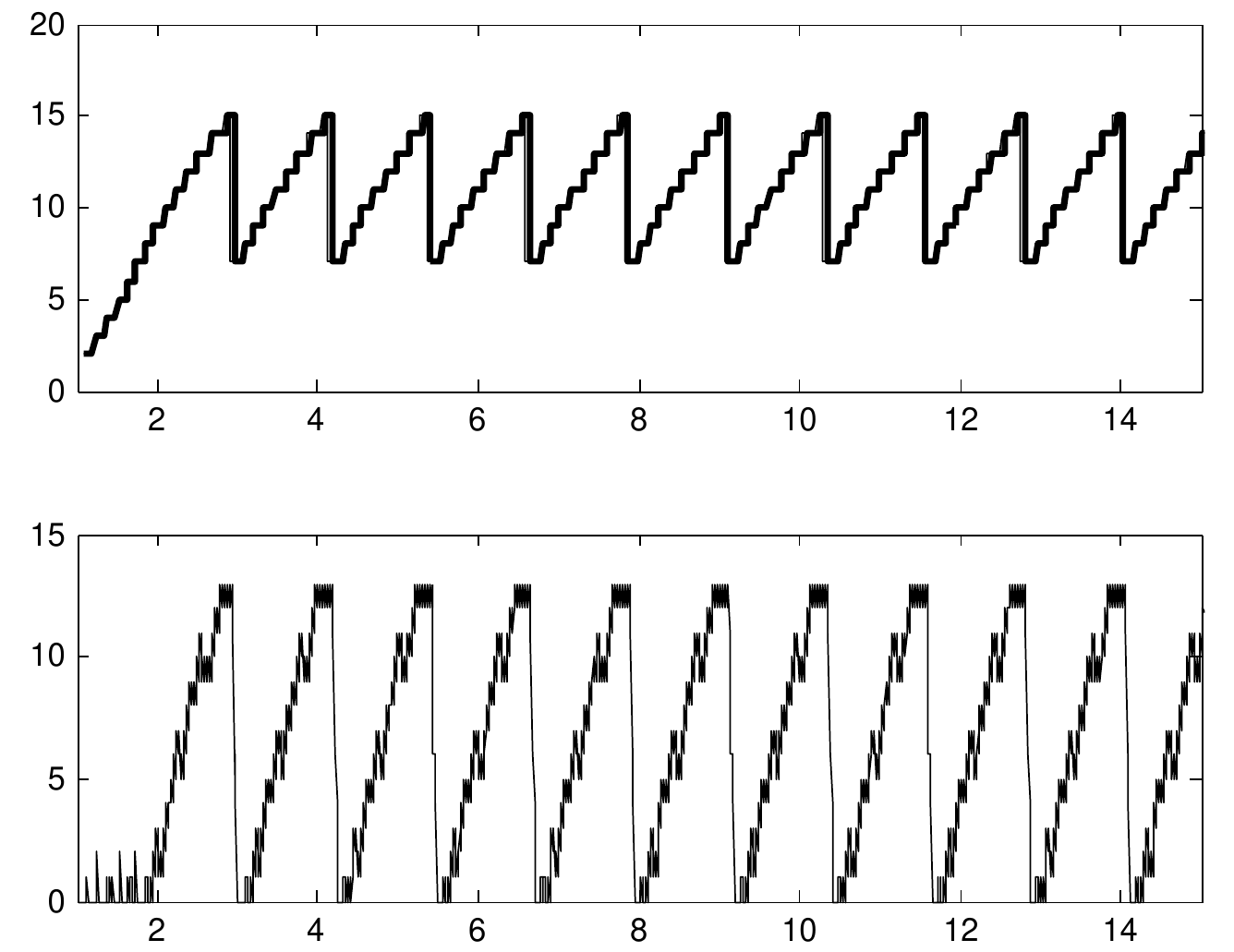}
                \caption{Congestion Window of the two users (up) and \\
		buffer queue size (bottom) in DropTail router}
                \label{fig:baseline_droptail}
        \end{subfigure}%
        ~ %add desired spacing between images, e. g. ~, \quad, \qquad etc.
        \begin{subfigure}[h!]{0.5\textwidth}
	   %\centering
                \includegraphics[scale=0.55,clip=true, trim=105 275 105 270]{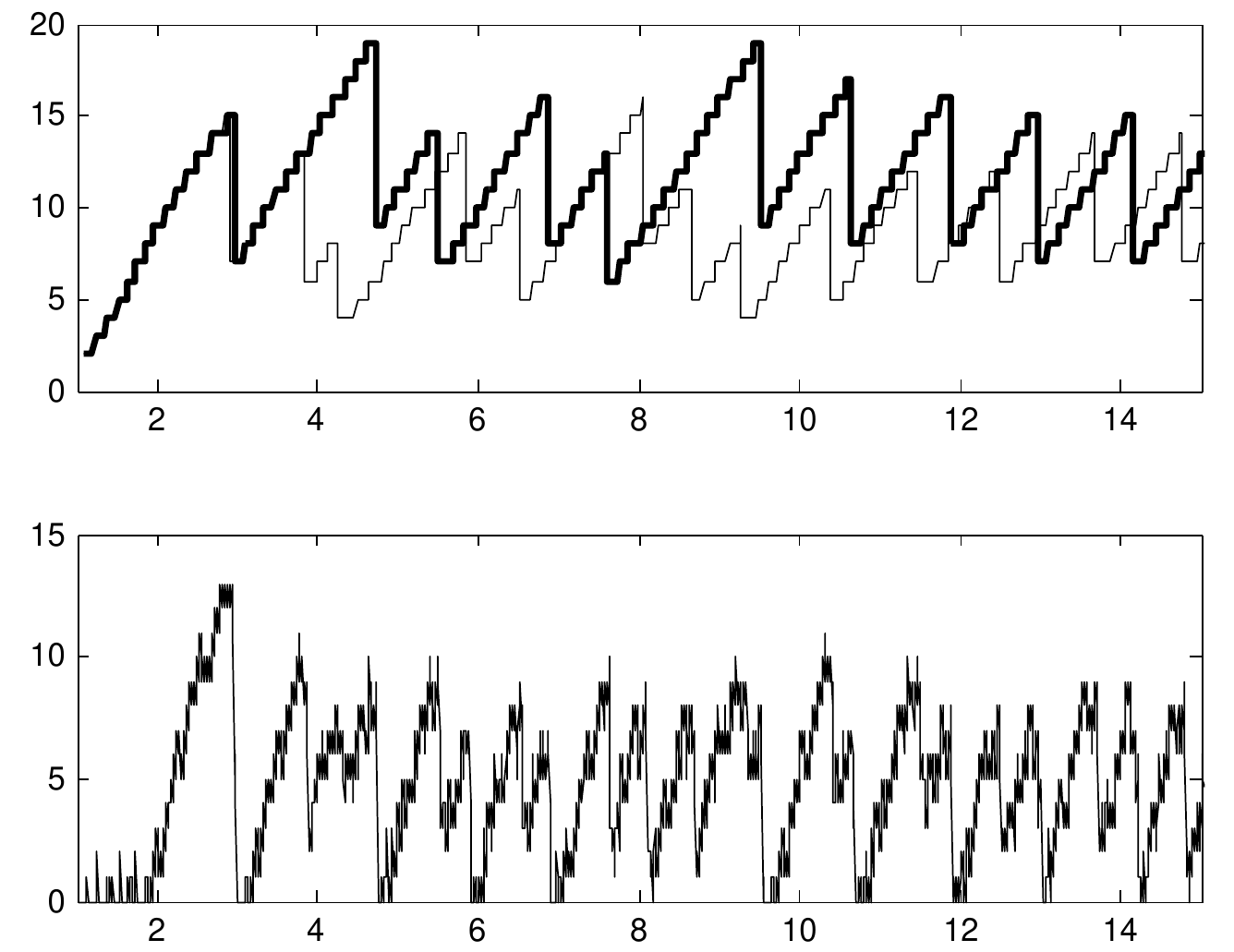}
                \caption{Congestion Window of the two users (up) and \\
		buffer queue size (bottom) in RED router}
                \label{fig:baseline_red}
        \end{subfigure}

	\begin{subfigure}[h!]{0.5\textwidth}
  	  \centering
  	     \includegraphics[scale=0.58,clip=true, trim=105 270 105 240]{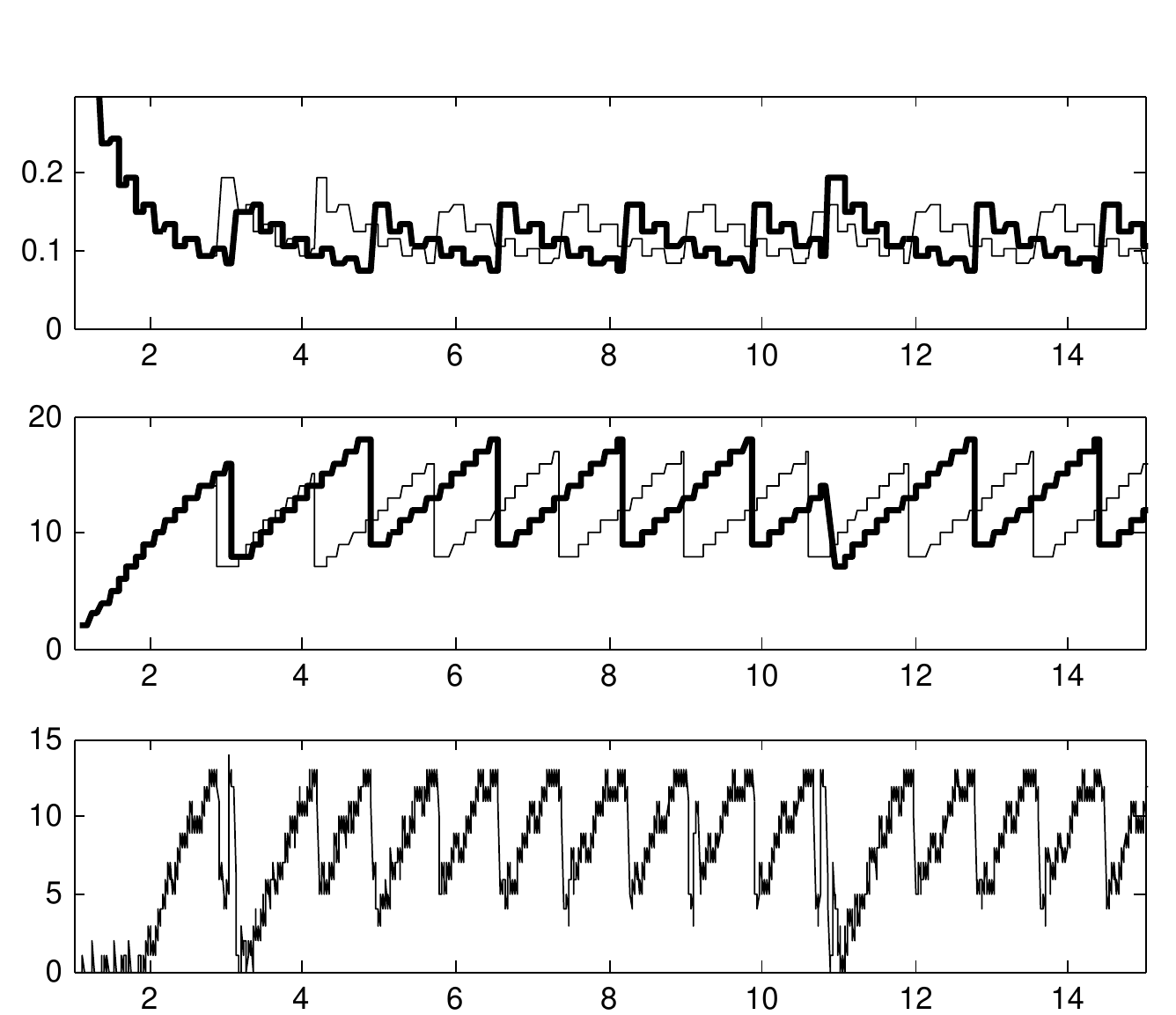}
  	  \caption{Index values of the two users (up), congestion windows of the two users (middle) and buffer queue size (bottom) in index-policy router}
  	  \label{fig:baseline_alpha1}
	\end{subfigure}
	\caption{Scenario 0 (Baseline): Simulation of a bottleneck with two equal users with standard TCP ($ \gamma_{ 1 },\gamma_{ 2 } = 0.5 $)}
\end{figure}
\label{baseline}
As a baseline scenario, we consider two symmetric
(equal) users that are sending data to a server through a bottleneck
router. Each user $ k = 1, 2 $ is halving her congestion window, i.e., $
\gamma_{ k } = 1/2 $.

The buffer size is set to the Bandwidth-Delay product of a single user,
\begin{align*}
\frac{1500 \cdot 10^{ 3 } \text{b/s} \cdot 2 \cdot (10^{-2} \text{s} + 10^{-2} \text{s})}{576 \text{B} \cdot 8 \text{b/B}}\approx 13.
\end{align*}

We present the evolution of the congestion window and the size of
the queue of the router in time for the DropTail case in
\autoref{fig:baseline_droptail}. As expected due to the well-known
phenomenon, the two users are completely synchronized.

We depict the evolution of the congestion window and the size of
the queue of RED policy in \autoref{fig:baseline_red}. We observe that
the users get unsynchronized and in this instance a periodic behavior
of the congestion window is not achieved because the packets are dropped randomly.

We show the evolution of the indices, the congestion window and the
size of the queue of the router in time for the index policy in
\autoref{fig:baseline_alpha1}. Interestingly, we can observe that
 users become ideally unsynchronized and as a consequence
 the utilization of the buffer increases. As it can be seen from
\autoref{results}, the throughput increases by $1.1\%$ and $0.6\%$
with respect to Droptail and RED policies. However, the
user fairness remains essentially the same as in
DropTail policy, but it improves user fairness comparing with RED.

\subsection{Scenario 1: Reducing Buffer Size}

\begin{figure}[t!]
\centering
        \begin{subfigure}[h!]{0.5\textwidth}
	  % \centering
               % \includegraphics[scale=0.45,clip=true, trim=55 275 55 275]{figs/SmallQueueSize/droptail/droptail_SmallQueueSize_4.pdf}
                \includegraphics[scale=0.55,clip=true, trim=105 275 105 275]{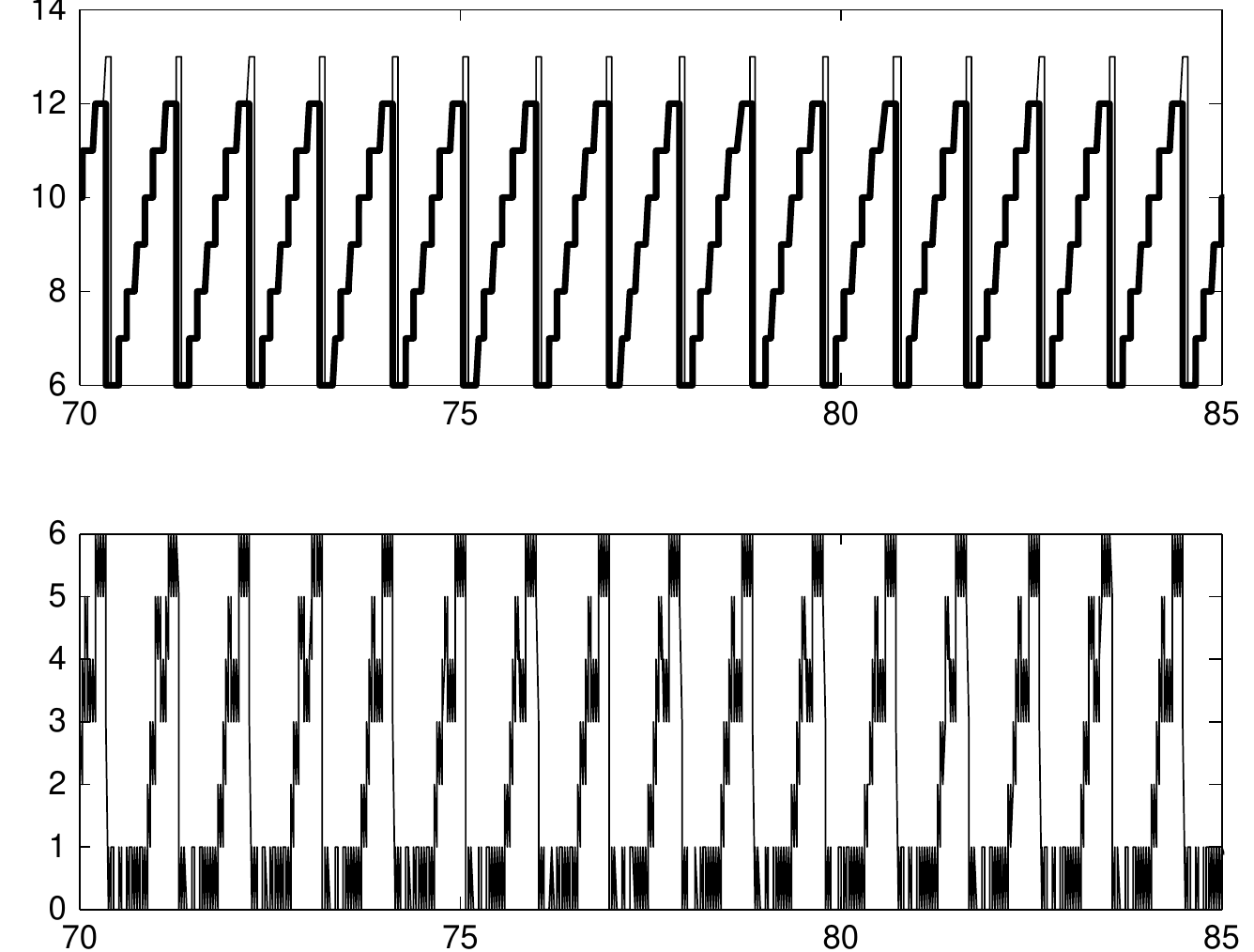}
                \caption{Congestion Window of the two users (up) and \\
		buffer queue size (bottom) in DropTail router}
                \label{fig:higher_buffer_droptail}
        \end{subfigure}%
        ~ %add desired spacing between images, e. g. ~, \quad, \qquad etc.
        \begin{subfigure}[h!]{0.5\textwidth}
	   %\centering
                \includegraphics[scale=0.52,clip=true, trim=105 265 105 265]{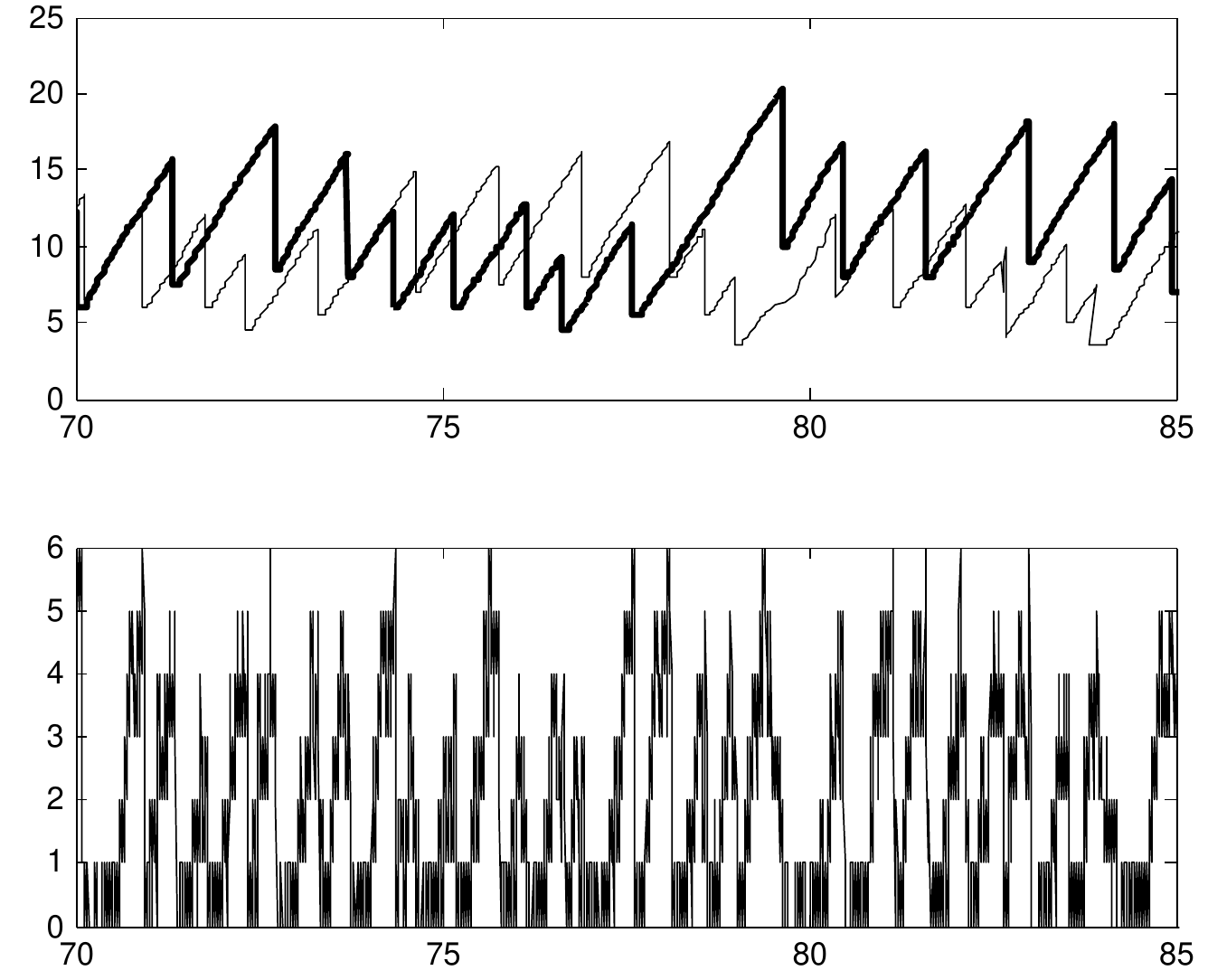}
                \caption{Congestion Window of the two users (up) and \\
		buffer queue size (bottom) in RED router}
                \label{fig:higher_buffer_red}
        \end{subfigure}

	\begin{subfigure}[h!]{0.5\textwidth}
  	  \centering
        	\includegraphics[scale=0.58,clip=true, trim=105 270 105 240]{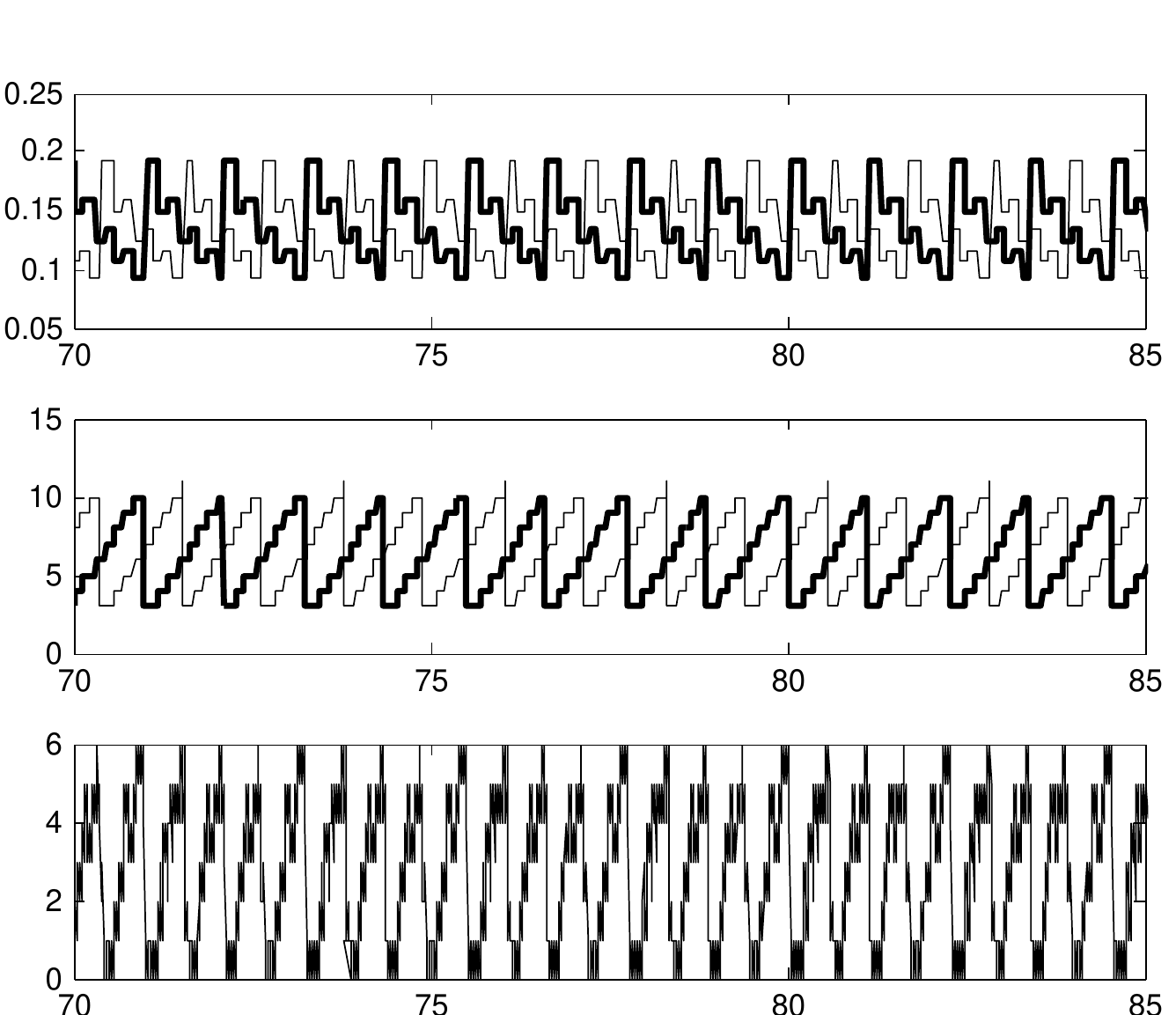}
  	   \caption{Index values of the two users (up), congestion windows of the two users (middle) and buffer queue size (bottom) in index-policy router}
    	   \label{fig:higher_buffer_alpha1}
	\end{subfigure}
	\caption{Scenario 1: Simulation of a bottleneck link with two equal users with standard TCP and a buffer size smaller than the bandwith-delay product}
\end{figure}

In this scenario we analyze the influence of setting a smaller buffer
size in the bottleneck queue for index policies, DropTail and RED.
To investigate this effect, we set the buffer size of the router to 6.

As we can see in \autoref{fig:higher_buffer_droptail}, with DropTail
users are synchronized and  the buffer is empty more time than in the
previous scenario.

In \autoref{fig:higher_buffer_red}, we observe again that the congestion
window with RED does not change periodicly.

In the case of the index policy (see
\autoref{fig:higher_buffer_alpha1}) the users are desynchronized,
and as a result the number of delivered packets is
larger than DropTail policy and also higher than RED.
We note that the fairness among users does not change much
in this case. We can also observe that the utilization
increases by $7.3\%$ and $6.7\%$ with respect to DropTail and RED,
but the RTT increases only $2.3\%$ and $2\%$, respectively.

The main conclusion of this scenario is that a buffer size smaller
than the bandwidth-delay product, the throughput of index policies
is larger than DropTail and RED.

\subsection{Changing Multiplicative Decrease Factor}

We illustrate the inter-protocol properties of the
index policy. In Scenario 2 we change the multiplicative decrease
factor of user 1 to $\gamma_1 = 0$, and in Scenario 3 to $\gamma_1=0.9$.

\subsubsection{Scenario 2:}

In this setting user 1 is conservative comparing to user 2, and
reinitializes the congestion window to 1  every time  a packet is
lost, i.e, $\gamma_1=0$.

\begin{figure}[t!]
\centering
        \begin{subfigure}[h!]{0.5\textwidth}
	  % \centering
                \includegraphics[scale=0.55,clip=true, trim=105 275 105 270]{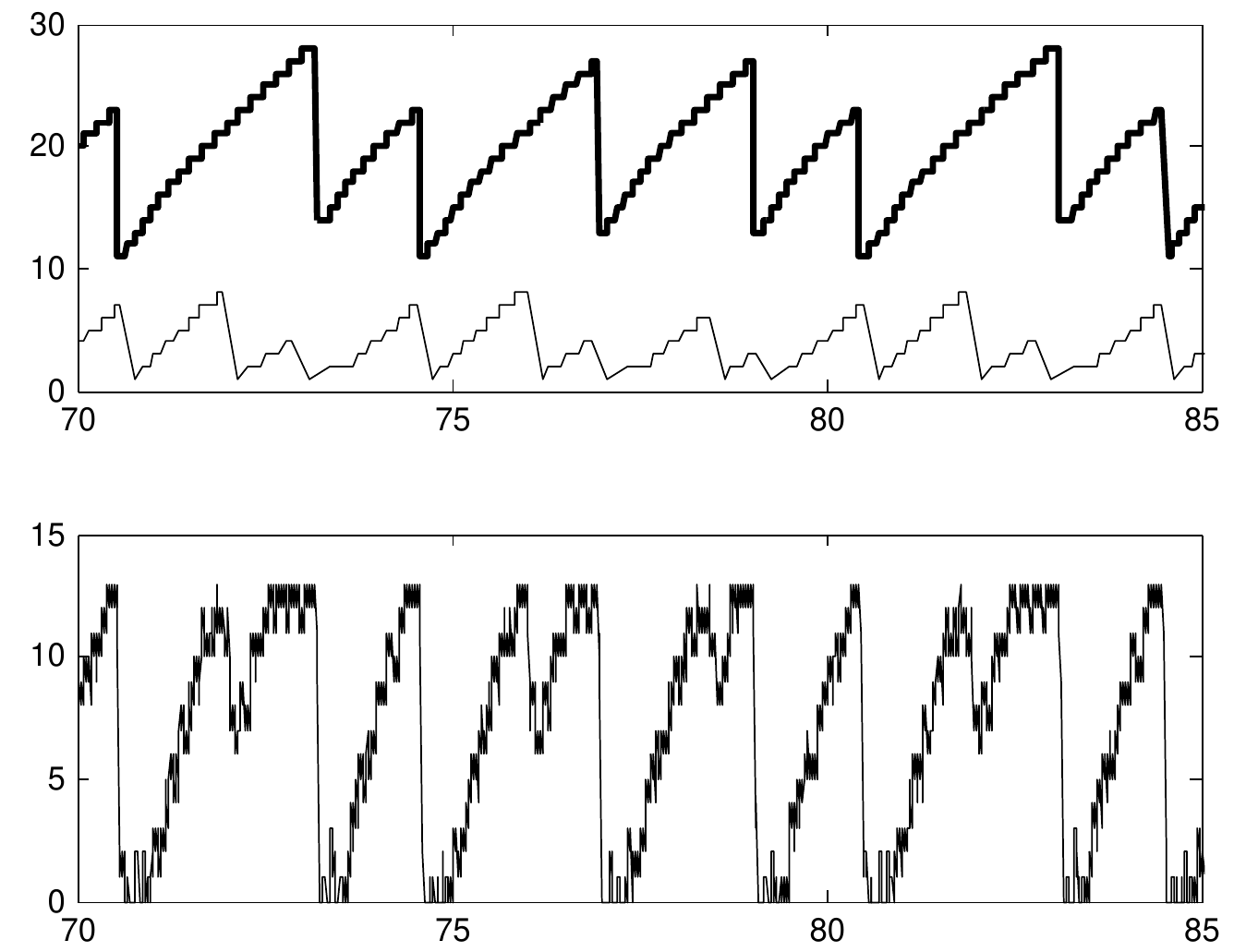}
                \caption{Congestion Window of the two users (up) and \\
		buffer queue size (bottom) in DropTail router}
                \label{fig:restarting_droptail}
        \end{subfigure}%
        ~ %add desired spacing between images, e. g. ~, \quad, \qquad etc.
        \begin{subfigure}[h!]{0.5\textwidth}
	   %\centering
                \includegraphics[scale=0.55,clip=true, trim=105 275 105 270]{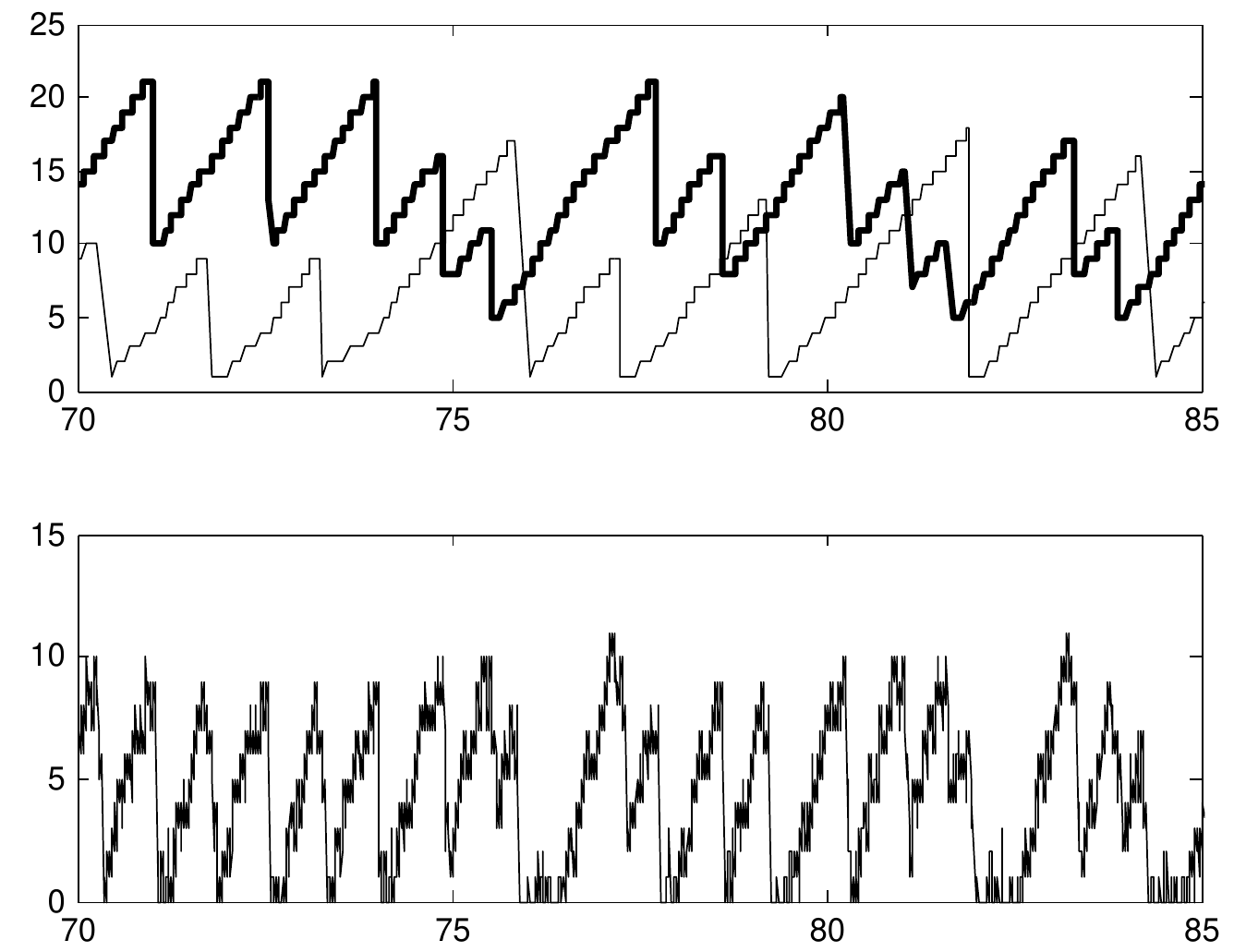}
                \caption{Congestion Window of the two users (up) and \\
		buffer queue size (bottom) in RED router}
                \label{fig:restarting_red}
        \end{subfigure}

	\begin{subfigure}[h!]{0.5\textwidth}
  	  \centering
     		\includegraphics[scale=0.58,clip=true, trim=105 275 105 240]{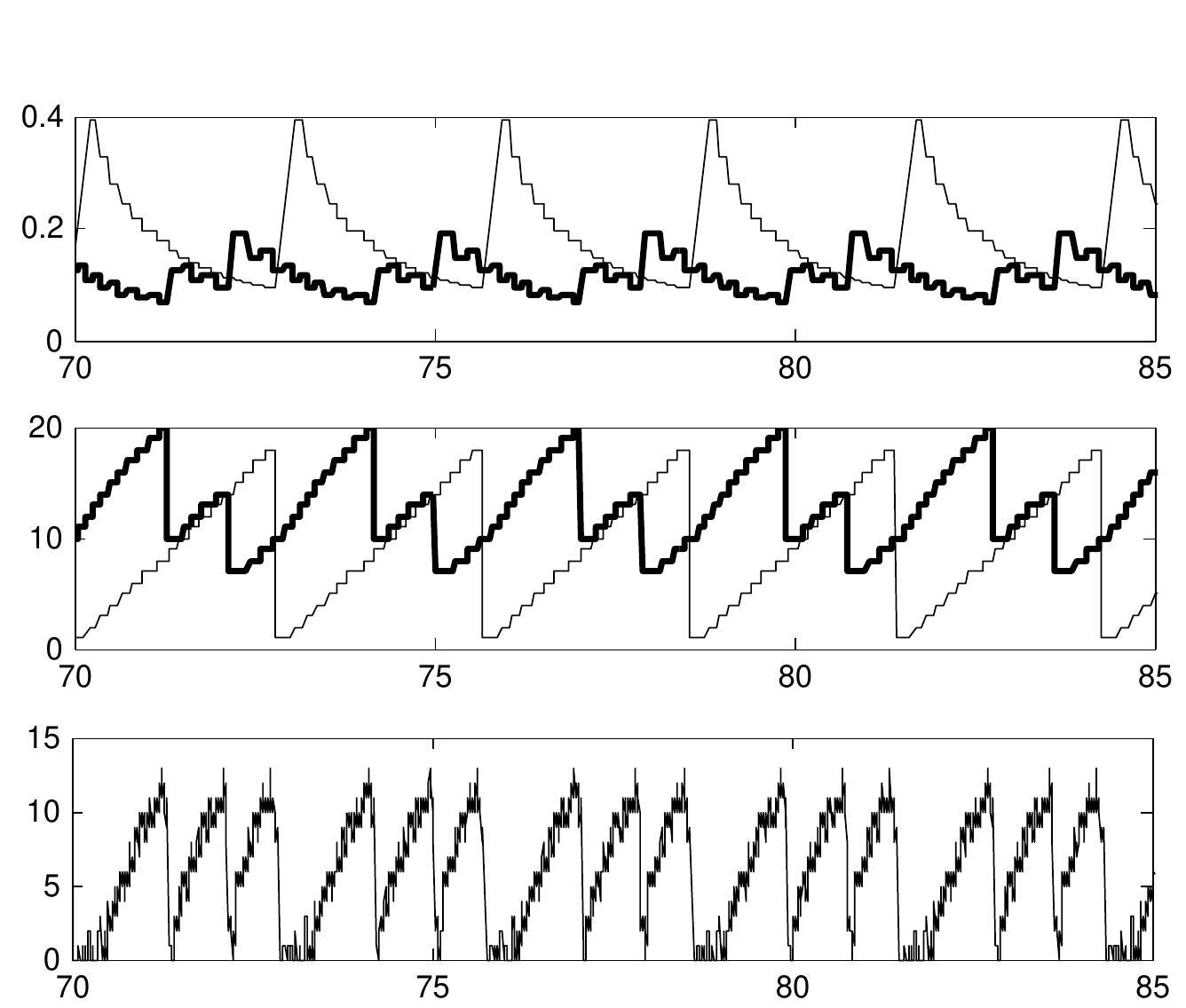}
  	\caption{Index values of the two users (up), congestion windows of the two users (middle) and buffer queue size (bottom) in index-policy router}
    	   \label{fig:restarting_alpha1}
	\end{subfigure}
	\caption{Scenario 2: Simulation of a bottleneck link with two users, where User1 has restarting TCP ($ \gamma_{ 1 } = 0 $) and User2 has standard TCP ($ \gamma_{ 2 } = 0.5 $)}
\end{figure}

We observe in \autoref{fig:restarting_droptail} that with DropTail
the congestion window of user 2 is consistently bigger. This implies
that the number of delivered packets by user 2 is much larger, as can be seen in
\autoref{results}.

In \autoref{fig:restarting_red} we see that once again RED
causes an unsynchronized behavior of the users.
Besides, RED policy improves fairness among users with respect to DropTail,
as we show in \autoref{results}.

With the index policy the congestion window of user 1 is reduced less often than
user 2 (see \autoref{fig:restarting_alpha1}). In this case, users are completely desynchronized with
the property of higher Jain's fairness value, which improves with respect to
DropTail and RED by $22.3\%$ and $5.5\%$, respectively. At the same time,
the utilization of index policy is
larger, so that the total number of delivered packets is increased
by $4.5\%$ and $2.6\%$ comparing with DropTail and RED, while the RTT of one user
increases by $1.1\%$ and $6.8\%$, respectively.

From this scenario, we concude that index policy improves user fairness and
throughput for users with different TCP models.

\subsubsection{Scenario 3:}

\begin{figure}[t!]
\centering
        \begin{subfigure}[h!]{0.5\textwidth}
	   %\centering
                \includegraphics[scale=0.53,clip=true, trim=105 265 105 265]{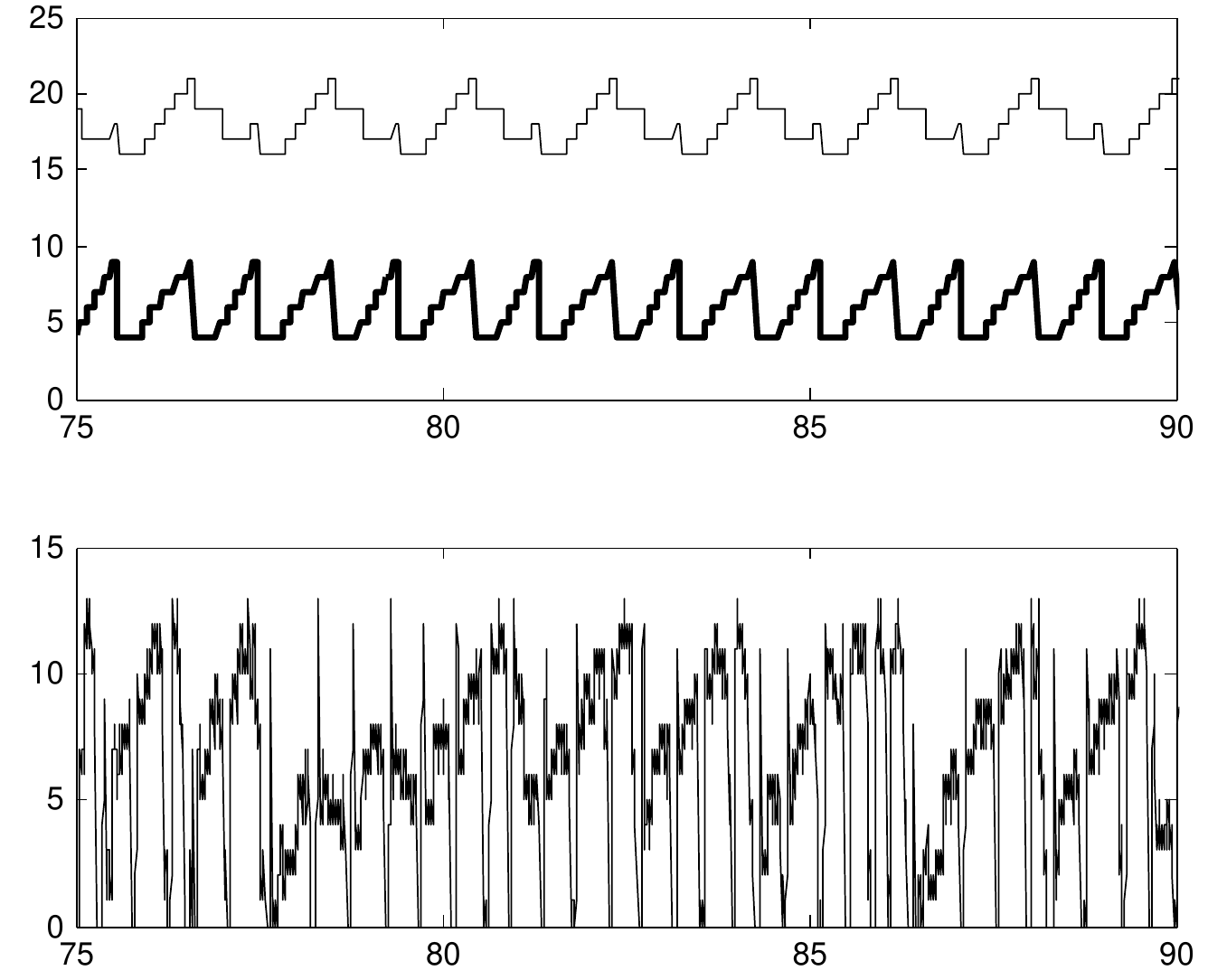}
                \caption{Congestion Window of the two users (up) and \\
		buffer queue size (bottom) in DropTail router}
                \label{fig:gamma09_droptail}
        \end{subfigure}%
        ~ %add desired spacing between images, e. g. ~, \quad, \qquad etc.
        \begin{subfigure}[h!]{0.5\textwidth}
	  % \centering
                \includegraphics[scale=0.55,clip=true, trim=105 275 105 265]{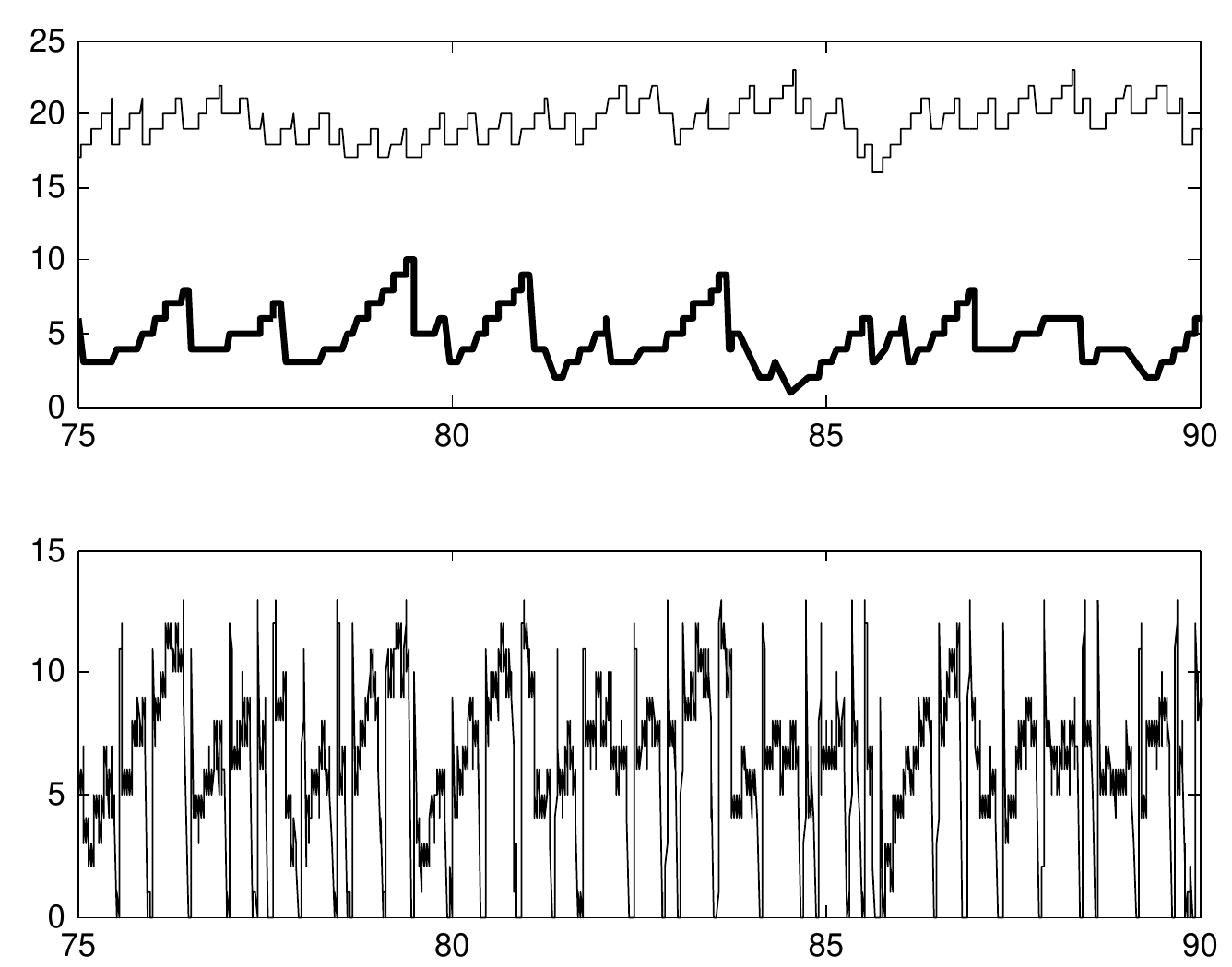}
                \caption{Congestion Window of the two users (up) and \\
		buffer queue size (bottom) in RED router}
                \label{fig:gamma09_red}
        \end{subfigure}

	\begin{subfigure}[h!]{0.5\textwidth}
  	\centering
        	\includegraphics[scale=0.58,clip=true, trim=105 275 105 240]{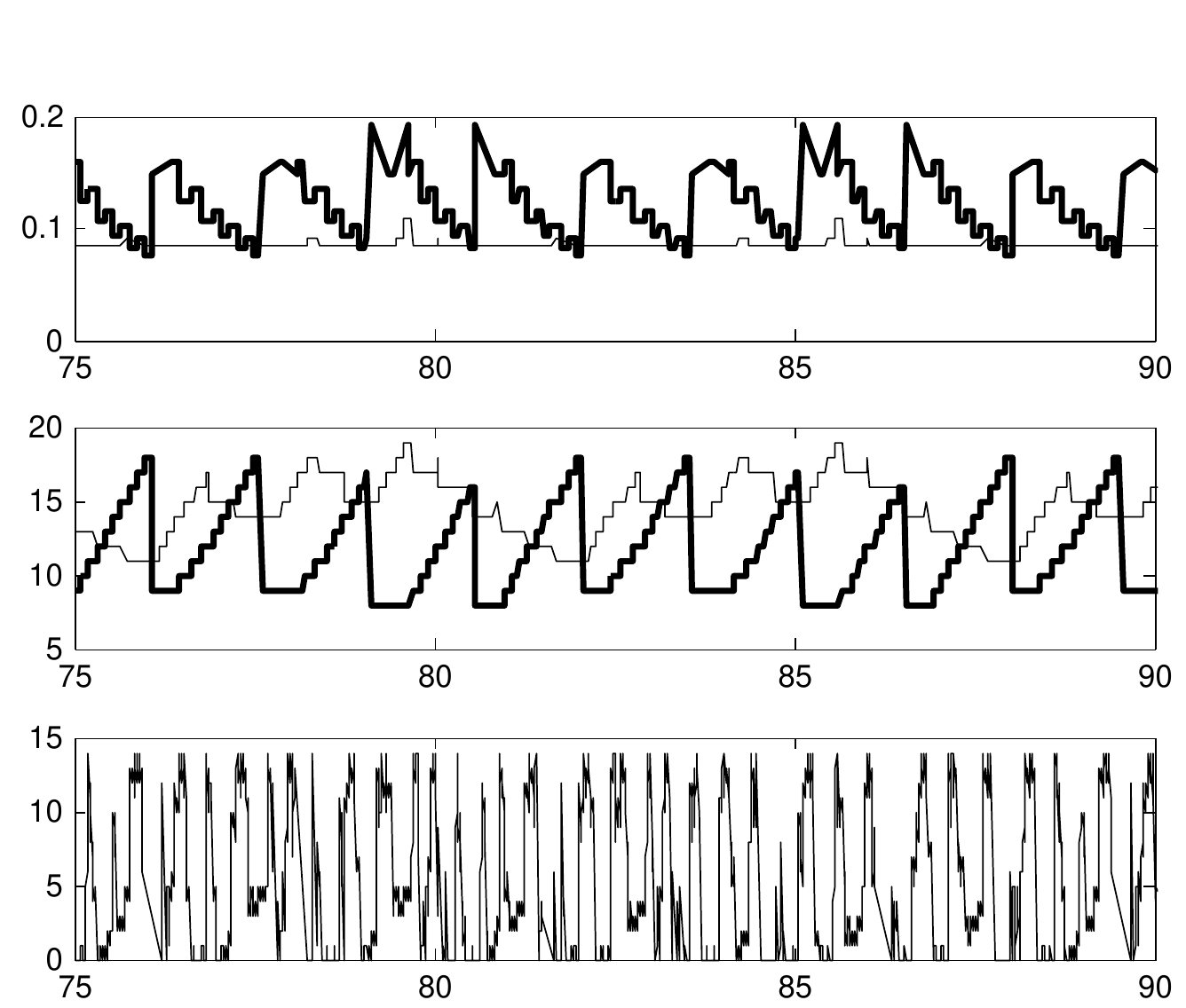}
  	\caption{Index values of the two users (up), congestion windows of the two users (middle) and buffer queue size (bottom) in index-policy router}
    	\label{fig:gamma09_alpha1}
	\end{subfigure}
	\caption{Scenario 3: Simulation of a bottleneck link with two users, where User1 has a TCP  with $ \gamma_{ 1 } = 0.9 $ and User2 has standard TCP ($ \gamma_{ 2 } = 0.5 $)}
\end{figure}

User 1 is now much more aggressive than user 2, since it barely
reduces its congestion window in response to a congestion event,
i.e. $\gamma_1=0.9$.

With DropTail we observe that user 1 has a significantly bigger
$cwnd$ in every moment (see \autoref{fig:gamma09_droptail}). This
illustrates that  DropTail is not fair in this setting.

In \autoref{fig:gamma09_red} we see that RED policy is not fair either,
because the congestion window of the user 1 is always higher
than the congestion window of user 2.

On the other hand, we see in \autoref{fig:gamma09_alpha1} that with the index
policy the difference in the congestion window between both users is
not big, which results in a significantly better Jain's
fairness index value, which is increased by $24.5\%$ and $11.6\%$
comparing with DropTail and RED, respectively. At the same
time, the utilization is increased by $10.5\%$ and
$2.8\%$ with respect to DropTail and RED and the RTT of one user
increases by $3.6\%$ and $5.2\%$.

The main contribution of this scenario consists on showing that for index policy the user
fairness is larger than DropTail and RED when a user is more agressive than the other.

\subsection{Scenario 4: Changing Propagation Delay}
\label{subsec:changing_prop_del}

\begin{figure}[t!]
\centering
        \begin{subfigure}[t]{0.5\textwidth}
	   %\centering
                \includegraphics[scale=0.55,clip=true, trim=105 275 105 265]{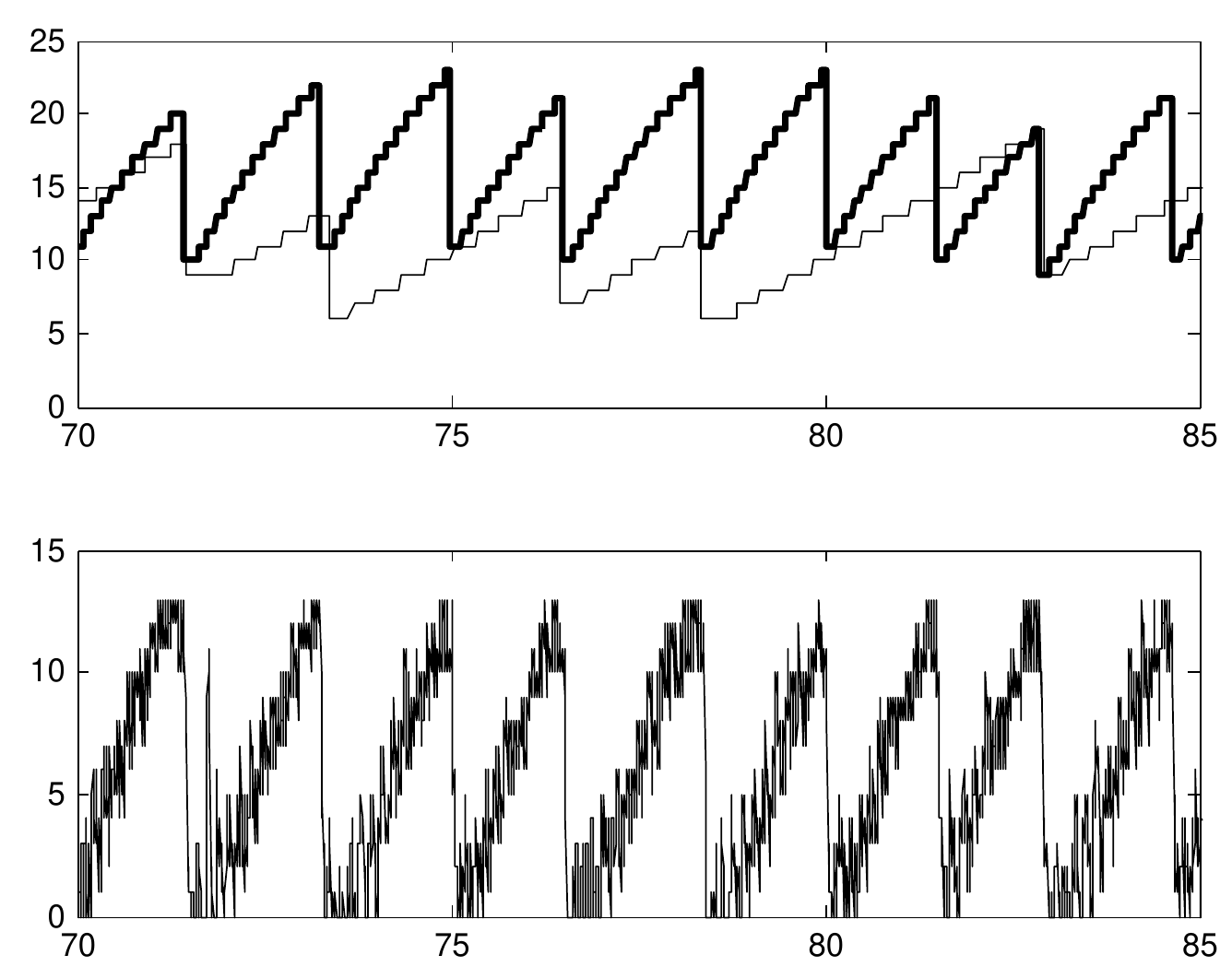}
                \caption{Congestion Window of the two users (up) and \\
		buffer queue size (bottom) in DropTail router}
                \label{fig:doubleRTT_droptail}
        \end{subfigure}%
        ~ %add desired spacing between images, e. g. ~, \quad, \qquad etc.
        \begin{subfigure}[t]{0.5\textwidth}
	   %\centering
                \includegraphics[scale=0.55,clip=true, trim=105 275 105 265]{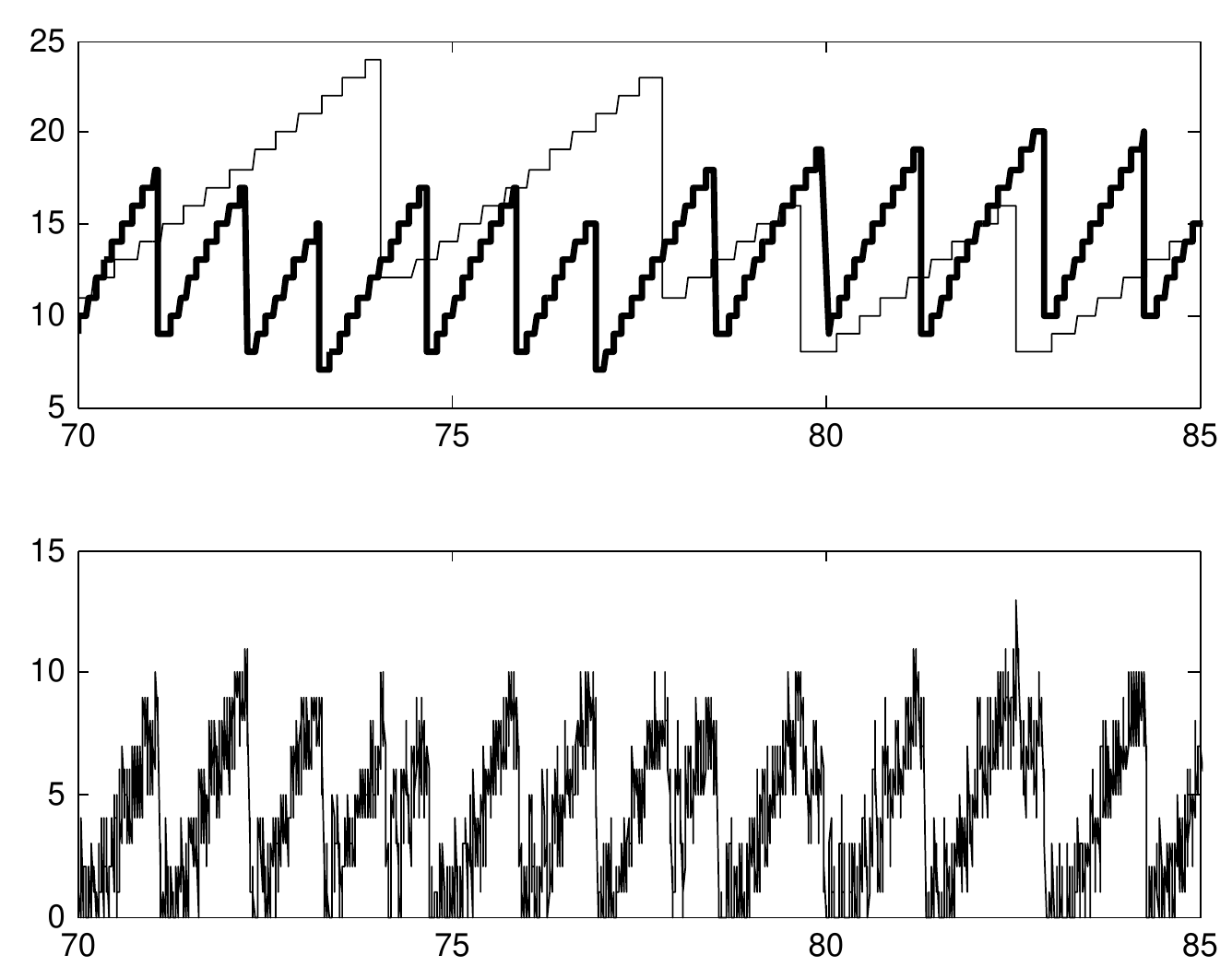}
                \caption{Congestion Window of the two users (up) and \\
		buffer queue size (bottom) in RED router}
                \label{fig:doubleRTT_red}
        \end{subfigure}

	\begin{subfigure}[h!]{0.5\textwidth}
  	   \centering
         	\includegraphics[scale=0.58,clip=true, trim=55 275 35 240]{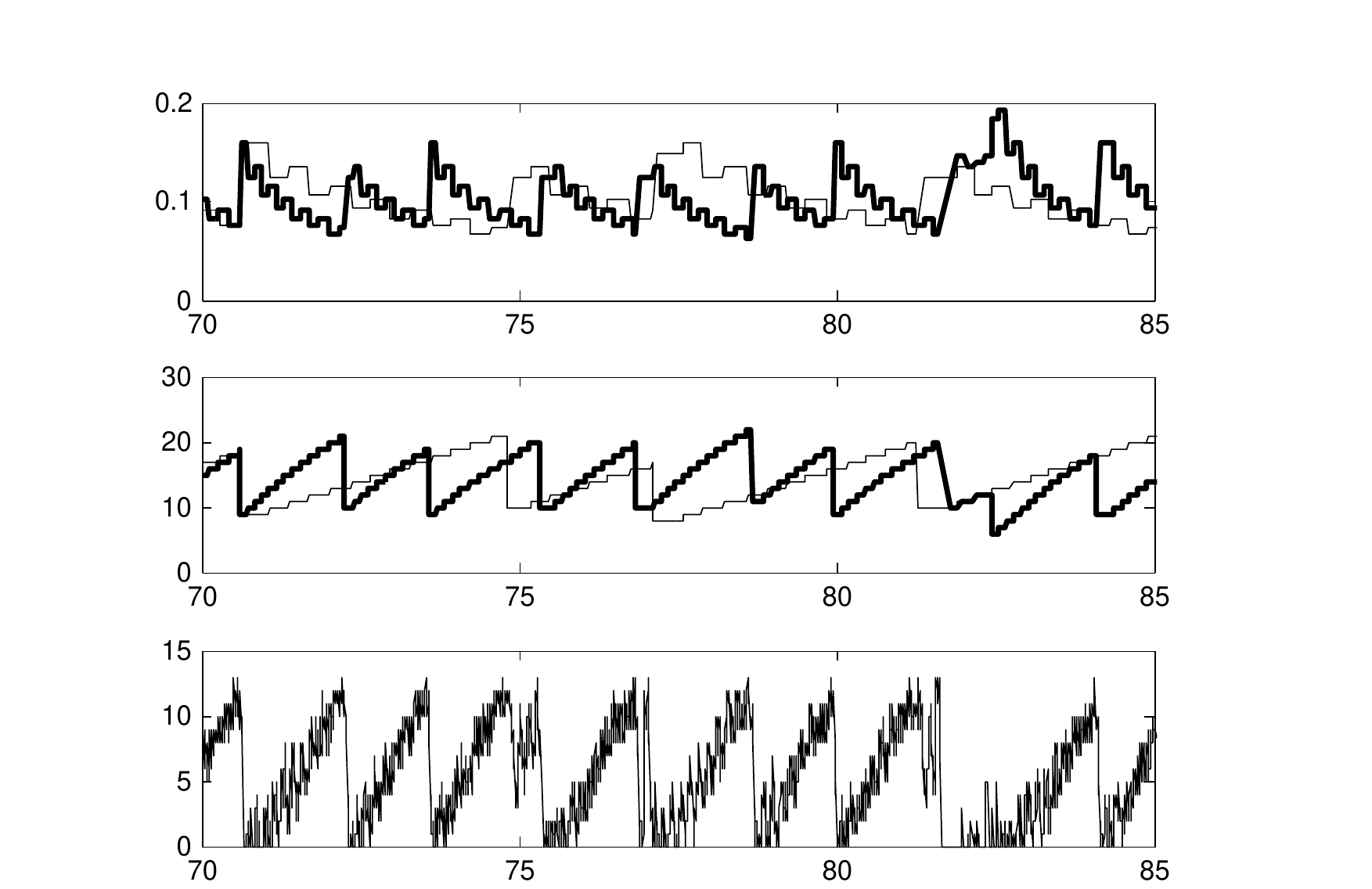}
	  \caption{Index values of the two users (up), congestion windows of the two users (middle) and buffer queue size (bottom) in index-policy router}
	    \label{fig:doubleRTT_alpha1}
	\end{subfigure}
	\caption{Scenario 4: Simulation of a bottleneck link with two users, where the propagation delay of User1 is $ 50 $ms}
\end{figure}

In this scenario we modify the propagation delay of the access link
of user 1 and we set it to $ 50 $ms.

In this setting, the congestion window
of both users under DropTail, RED and the index policy becomes very
similar. The main difference is that with the index policy, the users
gets again "ideally" unsynchronized, while the behavior of the congestion
window in DropTail and RED is more random. As we can see in the
Jain's index in \autoref{results}, the user fairness for
index policy is $24.5\%$ and $11.6\%$, larger than DropTail and RED
policies, respectively.

We explain the differences of the RTT of each user in DropTail, RED and
index policy. The values of the RTT of the user 1 for DropTail, RED and
index policy are $ 138.8 $ms, $ 137.8 $ms and $ 141 $ms, respectively.
For this user, the increasing of the RTT in index policy with respect to DropTail
and RED is $ 1.5\% $ and $ 2.3\% $.
On the oher hand, the values of the round-trip-time of the user 2 for
Droptail, RED and index policy are $ 58.8 $ms , $ 57.8 $ms and $ 61 $ms
respectively, so the increase of the RTT of this user with index policy
with respect to DropTail and RED is $3.7\%$ and $5.5\%$.

According to the results obtained in this scenario, we conclude that index policy
has the property of RTT fairness and higher throughput than DropTail
and RED policies.

\section{Discussion}
\label{sec:discussion}

\subsection{Cooperative Sources}

The ideal implementation of the proposed index policy is that the end-users (sources) are cooperative, as the TCP sources are
today, and they write the index in the packet header of every packet sent. In this way, there is no computational overhead,
because the index values can be computed offline for the TCP variant the source has implemented, and the indices can be stored
in a lookup table, where an index is assigned to every possible value of congestion window. This table needs to be accessed
only when (the integer part of) the congestion window changes, which happens approximately once per RTT, which is enough to
access the data. While (the integer part of) the congestion window does not change, all outgoing packets have the same index,
so the lookup table does not need to be accessed.

From the point of view of the router, an efficient implementation seems to be to maintain a dynamic table of indices of
all the buffered packets, with the pointer to the packet of lowest index (in FIFO order). That is, upon arrival of every
packet, its index is memorized, and upon departure of every packet, it is removed from the list. This dynamic table of indices
can thus be maintained separately from the regular buffer operation. When there is a requirement to drop/mark a packet, the
pointer indicates which packet it should be applied to. The index corresponding to the pointed packet can be interpreted as the
\emph{congestion price} (the higher the congestion price, the more congested the router is), which changes dynamically as the
buffer utilization evolves. Note further that such an implementation implies that the router \emph{does not need to identify
flows}.

In the following paragraphs we discuss several important features not explicitly covered by our model.

\paragraph{Short flows}
Our model assumes that the source has always data to transmit, so that the connection is permanent. This is a common assumption
in the TCP literature, however, in today practice the flows are often short, intertwined by long periods without transmission.
It is straightforward to extend our model to such flows: if the exact flow size is known by the source, then the MDP model will
have a two-dimensional state (including the remaining number of packets alongside the congestion window); if (more likely) the
exact flow size is unknown, e.g., because of the user impatience or incomplete information, it could be modeled for instance by
incorporating discounting. In either case, the indices are computed offline and the implementation would not be altered.

\paragraph{Timeouts and slow-start}
Our model can be extended in order to include the slow-start mode of the TCP algorithm. A second Markov chain representing the
slow-start mode must be considered so that we can interchange from one chain to the other, when the dynamics of the flow change
from slow-start mode to congestion avoidance mode and viceversa. Although we are able to generalize the model to include
slow-start, we leave it out of the paper, as we believe it would not affect the index values significantly.

\paragraph{Many-flows scalability}
The index policy requires packet header inspection, where the index value is read apart from the data (such as destination)
currently read. Implemented naively by reading before buffering, this may cause delays and decrease the input capacity of the
router. However, it may be perfectly acceptable to read the index once the packet has been buffered, thus avoiding any delays
in processing.

Another possibility is to read indices of only some of the packets (\emph{the power of few}), which is likely to identify a
packet with sufficiently low index (though maybe not the lowest and maybe not following the FIFO tie-breaking, which may not be
essential).

\paragraph{Multiple links}
In this paper we consider a single-bottleneck network, i.e., other routers on the path are implicitly assumed congestion-free.
We believe that this assumption is reasonable because the congestion of a flow when it traverses multiple links typically
occurs only in one router, often at the edge or at the server (see \cite{MW00} for details). Moreover, from
the point of view of the source that computes the index, it is indistinguishable where exactly the congestion occurs or whether
there are multiple bottlenecks on the route.

From the point of view of multiple-bottleneck routers, the situation is slightly different. If a given router believes that
further downstream there is a more congested router, it may be worth to drop/mark a packet earlier than usual (e.g., set the
buffer size smaller or increase the dropping/marking probability in AQM). Another possibility is to carry this information further
downstream by rewriting the index of the packet: decreasing the index by the current congestion price at every router on the path. Thus, the
packet is likely to traverse the whole congested route without
congestion notification only if its index is high (i.e., it is
efficient to transmit it) relatively to the sum of the congestion
prices in all the routers on the path, and relatively to other
packets. Thus, the dropped/marked packet would be of the flow that
most significantly decreases the congestion in the network by reacting
to a congestion notification.

\paragraph{Interaction with IPSEC}
The index policy requires including the index value in the IP header of the packets. In the case that IPSEC encapsulation is utilized, a portion of the IP packet is, at least, encrypted. In case that only the payload of the IP packet is encrypted, then the routing will not be altered, and our method can be applied directly. However, if all the IP packet is encrypted, then the packet is encapsulated in a new IP header that will be read by the routers. In this scenario, the index could be written in the new IP header, so that our packet method in the routers is also applicable.

\paragraph{Gradual implemetation} There is no need to implement the
index mechanism in all the routers at the same time, and it is in
general more beneficial for the bottleneck routers (e.g., edge routers
and servers), since non-congested routers will have their congestion
price close to zero. Also it is not necessary that all the end-users
start to write indices in the packet headers at the same time - those
without index could be treated in the routers as today or defining a
fixed index, which would incentivize the end-users to adopt the index
mechanism.

\subsection{Router-Only Implementation}

The packets may not come with their index in the header, either because the TCP sources are not cooperative, or because
modification of TCP headers is not implementable, or because adding the index cell to the packet header of the current TCP
variants is not acceptable. Then, our mechanism could be implemented at the routers only. The router must be able to recognize which packet belongs to which flow, the current window and the TCP variant (future behavior) of each flow, and compute (an estimation of) the index for each flow.

In this case, a possibility would be that routers focus only on long TCP flows. Traffic measurements on the Internet have consistently showed that even though most TCP flows are small in size, the majority of data is carried by long TCP flows. One could then implement a priority scheme, see for example \cite{KOR04}, such that short TCP flows receive high priority. As a consequence, losses will only be suffered by long TCP flows. At any given time, the number of concurrent long TCP flows is rather small (see \cite{MW00} for details), which makes feasible the implementation of an index-based algorithm to classify the packets. We refer to \cite{KOR04} for a reference justifying how flow-based routers can be implemented on the Internet.

%
%\paragraph{Short flows}
%Using the idea introduced by \cite{KOR04}, the router gives preference to short flows in the queue. This means that short-lives
%flows are not going to be congested in the queue. Thus, our scheme aims to control only long flows where congestion
%avoidance is the most important mode.
%
%In
%cross-platform routers, low packet loss of short flows is assured. According to this result, we claim that we can give
%preference to short flows so that our scheme aims to control only long flows.
%
%
%It may not be necessary to compute the index for every flow, e.g., those with currently higher congestion window are more
%likely to have lower index. The flows could also be aggregated, e.g., by input ports, among which the router decides according
%to indices, but within which a standard dropping/marking mechanism is implemented.
%
%However, since we focus only on the congestion of long-time flows, we state that it is enough to inspect some of the incoming packets (one out of every
%five packets).
%
%However, we state that our solution is scalable using the idea introduced by \cite{KOR04}.
%

\section{Conclusion}
\label{sec:conclusion}
In this paper we have introduced a rigorous framework of Markov
Decision Processes for the problem of optimal queue management in a
bottleneck router in order to achieve fast and fair transmission. We
have focused on showing tractability of the model and designed a
heuristic by means of transmission indices that can be implemented
for packet-level admission to the buffer.

The main goal has been to study in a benchmark topology the
fundamental features and performance of the proposed heuristic for
flows that behave according to TCP/IP protocol. A noticeable feature
is that the proposed heuristic manages to desynchronize (and even
counter-synchronize) the flows, so that network resources can be
used more efficiently. We have shown in NS-3 simulations that the
proposed heuristic significantly improves over the DropTail and RED policies
in several aspects, including fairness across users with different
TCP variants and fairness with respect to
different Round Trip Times. In scenarios where DropTail and RED performed
well in these types of fairness, the proposed heuristic maintained
the same levels of fairness, and improved the throughput at the same
time. We also believe that we provide fundamental framework and ideas
useful or directly applicable also in next-generation Internet
architectures such as the Information-centric networking (ICN)
and wireless networks.

We have made several simplistic assumptions in the modeling of TCP.
However, we believe that our approach opens an interesting research
avenue to design admission control policies in Internet routers. In
order to fully validate our approach, future work must address some
of the limitations of our model, namely, the efficient computation
of index values, impact of misbehaving TCP sources, simulations
 in more realistic scenarios.

%\section*{References}
%\bibliographystyle{ijocv081}
\bibliographystyle{elsarticle-num}
\bibliography{tcp}

\begin{thebibliography}{10}

\bibitem{ns3}
Network simulator, ver.3, (ns-3), 2008.
\newblock http://www.nsnam.org/.

\bibitem{rfc2581}
M.~Allman, V.~Paxon, and W.~Stevens.
\newblock {TCP} congestion control.
\newblock RFC2581, April 2002.

\bibitem{AltmanEtal2008}
E.~Altman, K.~Avrachenkov, and A.~Garnaev.
\newblock Generalized $\alpha$-fair resource allocation in wireless networks.
\newblock In {\em Proceedings of the 47th IEEE Conference on Decision and
  Control}, pages 2414--2419, Cancun, Mexico, December 2008.

\bibitem{mama2012}
K.~Avrachenkov, U.~Ayesta, J.~Doncel, and P.~Jacko.
\newblock Optimal congestion control of tcp flows for internet routers.
\newblock In {\em Proceedings of MAMA 2012}, 2012.

\bibitem{AvrachenkovJacko2010tcp}
K.~Avrachenkov and P.~Jacko.
\newblock Index policies for congestion control of {TCP} flows.
\newblock In preparation, 2010.

\bibitem{CK74}
V.G. Cerf and R.E. Kahn.
\newblock A protocol for packet network intercommunication.
\newblock {\em IEEE Transactions on Communications}, 22(5):637--648, 1974.

\bibitem{DukkipatiEtal2005}
N.~Dukkipati, M.~Kobayashi, R.~Zhang-Shen, and N.~McKeown.
\newblock Processor sharing flows in the {I}nternet.
\newblock In {\em Proceedings of IWQoS '05}, Lecture Notes in Computer Science
  3552, pages 271--285, 2005.

\bibitem{FF96}
K.~Fall and S.~Floyd.
\newblock Simulation-based comparisons of {T}ahoe, {R}eno and {SACK} {TCP}.
\newblock {\em ACM Computer Communication Review}, 26(3), 1996.

\bibitem{FJ93}
S.~Floyd and V.~Jacobson.
\newblock Random early detection gateways for congestion avoidance.
\newblock {\em IEEE/ACM Transactions on Networking}, 1(4):397--413, 1993.

\bibitem{GGW11}
J.C. Gittins, K.~Glazebrook, and R.~Weber.
\newblock {\em Multi-armed Bandit Allocation Indices}.
\newblock Wiley, 2011.

\bibitem{HRX08}
S.~Ha, I.~Rhee, and L.~Xu.
\newblock {CUBIC}: a new {TCP}-friendly high-speed {TCP} variant.
\newblock {\em Operating Systems Review}, 42(5):64--74, 2008.

\bibitem{Jacko2009met}
P.~Jacko.
\newblock Adaptive greedy rules for dynamic and stochastic resource capacity
  allocation problems.
\newblock {\em Medium for Econometric Applications}, 17(4):10--16, 2009.

\bibitem{Jacko2010jobs}
P.~Jacko.
\newblock Restless bandits approach to the job scheduling problem and its
  extensions.
\newblock In A.~B. Piunovskiy, editor, {\em Modern Trends in Controlled
  Stochastic Processes: Theory and Applications}, pages 248--267. Luniver
  Press, United Kingdom, 2010.

\bibitem{JackoSanso2012restarting}
P.~Jacko and B.~Sans\`{o}.
\newblock Optimal anticipative congestion control of flows with time-varying
  input stream.
\newblock {\em Performance Evaluation}, 69(2):86--101, 2012.

\bibitem{Jac88}
V.~Jacobson.
\newblock Congestion avoidance and control.
\newblock In {\em Proceedings of ACM SIGCOMM}, pages 314--329, August 1988.

\bibitem{JCH94}
R.~Jain, D.~Chiu, and W.~Hawe.
\newblock {A Quantitative Measure Of Fairness And Discrimination For Resource
  Allocation In Shared Computer Systems}.
\newblock {\em DEC Research Report TR-301}, 1994.

\bibitem{KatabiEtal2002}
D.~Katabi, M.~Handley, and C.~Rohrs.
\newblock Congestion control for high bandwidth-delay product networks.
\newblock In {\em Proceedings of SIGCOMM '02}, 2002.

\bibitem{Kel97}
F.P. Kelly.
\newblock Charging and rate control for elastic traffic.
\newblock {\em European Transactions on Telecommunications}, 8:33--37, 1997.

\bibitem{LestasEtal2008}
M.~Lestas, A.~Pitsillides, and P.~Ioannou.
\newblock Congestion control in computer networks.
\newblock In {\em Modeling and Control of Complex Systems}. CRC Press, 2008.

\bibitem{LLS05}
Y.~Li, D.J. Leith, and R.N. Shorten.
\newblock {Experimental Evaluation of TCP Protocols for High-Speed Networks}.
\newblock {\em IEEE/ACM Transaction on Networking}, 15(5):1109--1122, 2007.

\bibitem{MaEtal2008}
K.~Ma, R.~Mazumdar, and J.~Luo.
\newblock On the performance of primal/dual schemes for congestion control in
  networks with dynamic flows.
\newblock In {\em Proceedings of The 27th Conference on Computer Communications
  (IEEE INFOCOM)}, pages 960--967, 2008.

\bibitem{MW00}
J.~Mo and J.~Walrand.
\newblock Fair end-to-end window-based congestion control.
\newblock {\em IEEE/ACM Transactions on Networking}, 8(5):556--567, 2000.

\bibitem{Nino2007top}
J.~Ni\~no{-}Mora.
\newblock Dynamic priority allocation via restless bandit marginal productivity
  indices.
\newblock {\em TOP}, 15(2):161--198, 2007.

\bibitem{Puterman2005}
M.~L. Puterman.
\newblock {\em Markov Decision Processes: Discrete Stochastic Dynamic
  Programming}.
\newblock John Wiley \& Sons, Inc., Hoboken, New Jersey, 2005.

\bibitem{rfc3168}
K.K. Ramakrishnan, S.~Floyd, and D.~Black.
\newblock {The Addition of Explicit Congestion Notification {(ECN)} to IP}.
\newblock RFC3168, September 2001.

\bibitem{SS07}
S.~Shakkottai and R.~Srikant.
\newblock {\em Network optimization and control}.
\newblock NoW Publishers, 2007.

\bibitem{TSZS06}
K.~Tan, J.~Song, Q.~Zhang, and M.~Sridharan.
\newblock A {C}ompound {TCP} approach for high-speed and long distance
  networks.
\newblock In {\em Proceedings of IEEE INFOCOM}, 2006.

\bibitem{Vu07}
G.~Vu-Brugier, R.S. Stanojevic, J.~Leith, and R.N. Shorten.
\newblock A critique of recently proposed buffer-sizing strategies.
\newblock {\em ACM SIGCOMM Computer Communication Review}, 37(1):43--48, 2007.

\bibitem{WW90}
R.R. Weber and G.~Weiss.
\newblock On an index policy for restless bandits.
\newblock {\em Journal of Applied Probability}, (27):637--648, 1990.

\bibitem{WJLH06}
D.X. Wei, C.~Jin, S.H. Low, and S.~Hegde.
\newblock {{FAST TCP}: {M}otivation, architecture, Algorithms, Performance
  Experimental Evaluation of {TCP} Protocols for High-Speed Networks}.
\newblock {\em IEEE/ACM Transaction on Networking}, 14(6):1246--1259, 2006.

\bibitem{Whittle1988}
P.~Whittle.
\newblock Restless bandits: Activity allocation in a changing world.
\newblock {\em A Celebration of Applied Probability, J. Gani (Ed.), Journal of
  Applied Probability}, 25A:287--298, 1988.

\end{thebibliography}

\end{document}